\documentclass{jfm}
\usepackage{graphicx}
\usepackage{epstopdf, epsfig}
\usepackage{natbib}
\usepackage[colorlinks=true,citecolor=blue]{hyperref}
\usepackage{float}
\usepackage{tikz}
\usepackage{soul}
\usepackage{makecell}
\usepackage{tikz}
\usepackage{scalerel}
\usepackage{amsmath}

\usepackage{lineno}

\shorttitle{Freestream turbulence effects on wind turbine wakes}
\shortauthor{N. Biswas, and O.R.H. Buxton}
\title{Effect of freestream turbulence on the coherent dynamics of a wind turbine wake
}

\author{Neelakash Biswas\aff{1} \corresp{\email{n.biswas20@imperial.ac.uk}},
  Oliver R.H. Buxton \aff{1}
}

\affiliation{\aff{1}Department of Aeronautics, Imperial College London,
London, SW7 2AZ, UK

}

\graphicspath{{figures/}}

\newcommand{\rfA}[1]{{\textcolor{black}{#1}}} 
\newcommand{\rfB}[1]{{\textcolor{black}{#1}}} 
\newcommand{\rfC}[1]{{\textcolor{black}{#1}}} 
\newcommand{\rfD}[1]{{\textcolor{black}{#1}}} 
\newcommand{\rfE}[1]{{\textcolor{black}{#1}}} 

\begin{document}

\maketitle

\begin{abstract}
{The wake of a model wind turbine exposed to \rfB{incoming freestream turbulence (FST)} with a variety of turbulent characteristics is studied through Particle Image Velocimetry experiments. \rfC{The FST cases were produced using different passive turbulence generating grids. The cases spanned turbulent intensities ($T_i$) in the range $1.3\% \lesssim T_i\lesssim 14\%$ and only considered short integral length scales $L_v \lesssim 0.2D$ (where $D$ is the turbine diameter)}. Increasing $T_i$ and $L_v$ in this range resulted in an earlier breakdown of the tip vortices which in turn resulted in an earlier onset of wake recovery. \rfA{For all the FST cases considered, the initiation of wake meandering was found to be related to an intrinsic instability of the turbine, even for the cases with the highest FST levels. The amplitudes of wake meandering were similar for all the cases in the near wake ($x<2D$), but the amplitudes in the far wake ($x>4D$) were discernibly higher for all the FST cases compared to the no grid case (lowest $T_i$), primarily due to the early break down of the tip vortices.} Deeper insights into the origins, and subsequent evolution, of the various coherent motions (characterised by particular frequencies) in the presence of FST are obtained through analysis of the multi-scale triple-decomposed coherent kinetic energy budgets. The wake meandering modes in the presence of FST are shown to better utilize the mean velocity shear, extracting more energy from the mean flow while other sources such as non-linear triadic interactions and diffusion also become important. } 
\end{abstract}


\section{Introduction}

{Wind turbine wakes exhibit a large range of coherent structures such as the tip/root vortices, distinct vortex sheddings from the tower or the nacelle and wake meandering \citep{sherry2013interaction, okulov2014regular, de2021pod, biswas2024effect}. These coherent structures play a critical role in determining the spatio-temporal evolution of the wake of a turbine \citep{lignarolo2015tip, porte2020wind, biswas2024effect, biswas2024energy} and hence, can have large impacts on farm-level performance \citep{vermeer2003wind, barthelmie2007modelling, sanderse2011review, stevens2017flow}. The dynamics of the coherent structures depends on several factors including (a) the geometry of the turbine, such as the design of the blades \citep{abraham2023experimental, dong2023characteristics}, (b) the operating conditions, such as the tip speed ratio $\lambda$, (defined as $\lambda = \Omega R/U_{\infty}$, where $\Omega$ is the turbine's rotational speed, $R$ is the turbine radius and $U_{\infty}$ is the freestream velocity) \citep{sherry2013interaction, okulov2014regular,  biswas2024effect, biswas2024energy} and (c) freestream conditions such as the freestream turbulence (FST) level \citep{mao2018far, hodgson2023effects, bourhis2025impact}. }\\

Real wind turbines are almost always exposed to some level of FST. FST can be introduced for instance, due to the presence of the atmospheric boundary layer which is characterized by a broad range of scales due to the inherently large Reynolds number, or due to the presence of an upstream obstacle such as another wind turbine \citep{porte2020wind, kosovic2025impact}. The nature and the level of turbulence in an atmospheric boundary layer is linked to the boundary layer characteristics such as the boundary layer thickness and atmospheric stability \citep{pena2016ten, vahidi2024influence, kosovic2025impact}. The level of FST is often measured using the turbulence intensity ($T_i$), defined as $T_i = \sqrt{(\overline{u'^2})}/U_{\infty}$ (where $\sqrt{(\overline{u'^2})}$ is the standard deviation of the velocity time history). {In onshore sites wind turbines are exposed to higher $T_i$ ($T_i<15\%$) for most of their operating life \citep{elliott1990effects, wagner2011accounting, pena2016ten}, while for offshore sites freestream $T_i$ is generally milder: $6\%<T_i<8\%$ \citep{barthelmie2005ten, turk2010dependence}, {due to the reduced surface roughness}}.\\




{The effect of FST on the performance of a wind turbine is multifaceted.} FST can impart fluctuating loads \citep{eggers2003wind, stanislawski2023effect, robertson2019sensitivity, de2025influence} and significantly alter the power output from a turbine \citep{chamorro2013interaction, maganga2010experimental, blackmore2016effects}. Not only can it alter the performance of a standalone turbine, it can modify the spatio-temporal evolution of the wake, which can significantly change farm level performance \citep{porte2020wind}. On one hand, FST can make the wake meander more {in the lateral direction} \citep{espana2012wind, porte2020wind}, potentially reaching other turbines in a staggered arrangement, imparting harmful unsteady loads \citep{larsen2008wake}. On the other hand, FST can help {re-energise} the wake by enhancing {lateral/vertical momentum transport from the freestream} which can be beneficial for the downwind turbines in terms of {power production} \citep{gambuzza2023influence, vahidi2024influence}. \\

The effect of FST has been studied through a combination of numerical simulations (largely using large eddy simulation, LES), laboratory experiments using a small-scale representative wind turbine, and field experiments. High levels of FST have been shown to break down the tip vortices in the near wake \citep{chamorro2009wind, wu2012atmospheric}. {These vortices act as a `shield' in the near wake, inhibiting the exchange of mass, momentum and kinetic energy \citep{medici2005experimental}}. \citet{lignarolo2015tip} performed particle image velocimetry (PIV) experiments on a two-bladed wind turbine model and showed that there was a net kinetic energy entrainment into the wake only after the tip vortices broke down. \rfC{Similar conclusions were reached for a three-bladed wind turbine \citep{eriksen2017development, biswas2024energy}.} Therefore, high FST levels can accelerate wake recovery by breaking down this shield of vortices. In fact, a linear relationship between $T_i$ and the wake recovery rate has been reported \citep{niayifar2016analytical, carbajo2018wind, gambuzza2023influence}.\\

In the near wake, freestream turbulence breaks down the dominant coherent structures \textit{i.e.} the tip vortices. In contrast, freestream turbulence likely plays a different role on the dominant far wake structures, which have long been primarily associated with wake meandering. Although the dominance of wake meandering in the far wake has been known for many years, there still remain varied opinions on the genesis and characteristics of the meandering motion. For instance, a distinct wake meandering frequency has been reported when freestream turbulence intensity was negligible \citep{okulov2014regular, biswas2024effect} thereby associating wake meandering to a rotor scale bluff body shedding type instability \citep{medici2008measurements, howard2015statistics}. In contrast, in other works wake meandering has been treated as a `random' unsteady oscillation of the far wake driven by freestream turbulence \citep{porte2020wind}. {Here, the wind turbine wake was thought to be passively advected by the large scale structures present in the freestream turbulence \citep{larsen2008wake}. \rfC{However, later works suggested that wind turbines do not simply advect the structures present in the freestream, instead, a turbine has a dynamic response \textit{i.e.} certain frequencies in the freestream can be amplified in the wake \citep{chamorro2012evolution, singh2014homogenization, mao2018far,heisel2018spectral,gupta2019low, hodgson2023effects}, highlighting the importance of characterising the spectral content of the inflow. \citet{chamorro2012evolution} proposed that a turbine can be modeled as an `active filter', generating or amplifying certain frequencies while dampening others. Similar results were obtained by \citet{singh2014homogenization} who showed that the turbine made the velocity fluctuations in the wake more `homogenised' compared to the incoming flow.} 

\citet{mao2018far} identified nonlinear optimal perturbations to a wind turbine wake modeled as an actuator disk. \rfB{They reported that as the inflow perturbation amplitude increased, the optimal perturbation frequency reduced from {$St= 0.64$} to $St=0.25$ (where $S_t$ denotes the Strouhal number based on the turbine diameter and freestream velocity) for $Re=1000$ (based on turbine radius and freestream velocity).} As $Re$ was increased to 3000, the optimal perturbation frequency further reduced to $St=0.16$. These low frequencies fall well within the accepted range of wake meandering frequencies \citep{medici2008measurements, chamorro2013interaction, okulov2014regular}. Beyond $Re=3000$, \citet{mao2018far} noted that the optimal frequency did not change, indicating a possible $Re$ invariance {at all application specific Reynolds numbers}. \rfA{Similarly, the LES study of \citet{foti2018similarity} showed that the wake meandering amplitude, when properly scaled by the thrust, was largely invariant of the rotor size that spanned nearly 3 orders of magnitude (see figure 8(b) of \citet{foti2018similarity}). The authors however, treated wake meandering as more of a bluff body shedding type instability of the turbine \citep{howard2015statistics}. Therefore, there is now a general consensus that in real wind turbines exposed to FST, the nature and intensity of wake meandering can depend on multiple factors including the turbine geometry, operating condition and freestream turbulence \citep{howard2015statistics, mao2018far, gupta2019low, gambuzza2023influence, dong2023characteristics, biswas2024effect, wang2025effects}. The relative importance of each of these factors is still not fully understand and can well differ in the near wake and the far wake of the turbine \citep{foti2018similarity, dong2023characteristics, bourhis2025impact}. \citet{dong2023characteristics} studied three different blade designs under three different FST conditions but at similar $C_T$. Within $x<3D$, the wake meandering amplitudes (measured using the standard deviation of the wake centreline) were similar for the different blade designs and turbulence levels. Further downstream however, the blade design producing the strongest hub vortex showed significantly stronger wake meandering. For all the blade designs, in the far wake, the high FST case showed slightly higher wake meandering compared to low/medium FST cases (see figure 19 of \citep{dong2023characteristics}).}

\begin{figure}
  \centerline{
  \includegraphics[clip = true, trim = 0 0 0 0 ,width= 0.85\textwidth]{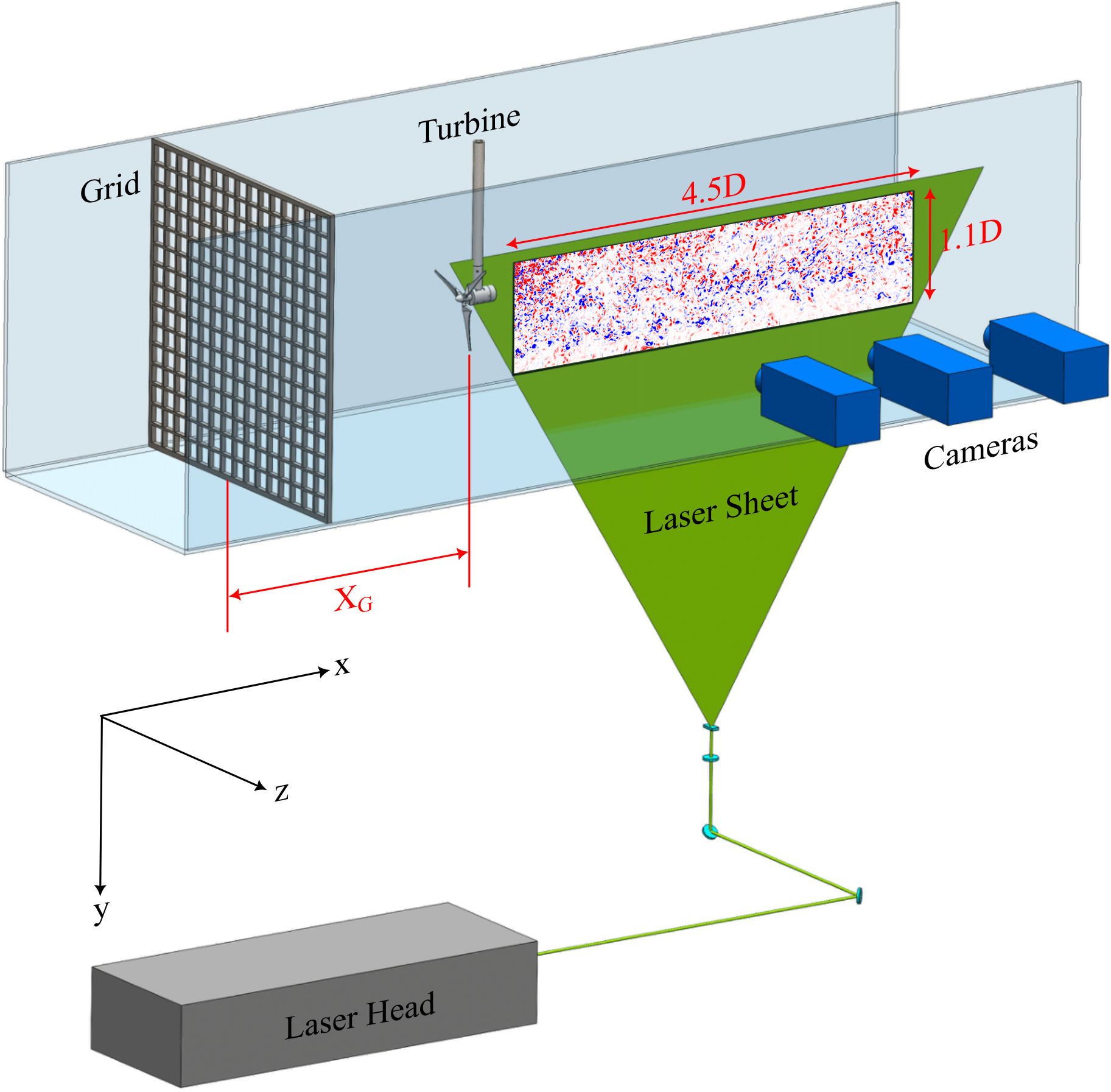}  }
 \caption{Schematic of the experimental setup. }
\label{fig:Schematic}
\end{figure}

\rfC{While the need to characterise the spectral content of the inflow has been realised for some time, the exact role of the freestream length/time scales on the dynamics of the wind turbine wake and the wake recovery process is still not fully understood. Larger freestream characteristic length scales were shown to induce a stronger wake meandering and were linked to a faster kinetic energy entrainment from the freestream towards the far wake \citep{blackmore2014influence, vahidi2024influence}. These studies however modeled the turbine as an actuator disk, hence the effect of the tip vortices in the near field was ignored. Interestingly, \citet{hodgson2023effects} reported a shorter near wake region and faster wake recovery in wind turbines exposed to a flow with shorter integral time scales. The authors modeled the wind turbine using the actuator line method (\textit{i.e.} the tip vortices were captured). Similarly, the PIV experiments by \citet{gambuzza2023influence} showed that for very large integral time scales in a non-Kolmogorov type flow, the wake recovered slowly compared to an inflow with a similar $T_i$ but shorter integral time scale. The authors linked this observation to the delayed break-up of the tip shear layer for this type of flows. Interestingly however, the recent experimental work by \citet{bourhis2025impact} did not find any clear influence of the freestream integral length scales on the strength of the leading tip vortex frequencies. These results indicate that the effect of the freestream characteristic length scales on the near and far wake dynamics of a turbine and the wake recovery process is still far from clear.}   \\

 \rfA{In this work, we explore answers to some of these appealing questions by performing a series of time-resolved PIV experiments on a wind turbine model exposed to \rfB{incoming freestream turbulence} having different turbulence intensities ($T_i$) and integral length scales ($L_v$). The wind turbine model included a nacelle and tower mimicking a real wind turbine. In particular, we attempt to better understand the following aspects: (a) How does the {energy content at} the dominant frequencies in the near-wake change with freestream turbulence? (b) In the absence of freestream turbulence, \citet{biswas2024effect} found a scaling between the wake meandering frequency and $\lambda$ indicating that wake meandering could be a rotor scale bluff body shedding type instability. How does the nature of wake meandering change in the presence of freestream turbulence? (c) How does freestream turbulence affect the energy exchanges between the various coherent modes \rfC{and how does it impact the wake recovery process}? (d) Finally, in all of the above what are the relative roles of $T_i$ and $L_v$?}

\section{Methodology}

Particle image velocimetry (PIV) experiments were conducted in the hydrodynamics flume in the Department of Aeronautics at Imperial College London. The flume had a cross-sectional area of $60 \times 60$ cm$^2$ at the operating water depth. \rfB{A schematic of the experimental setup is shown in figure \ref{fig:Schematic}.} The turbine diameter ($D$) was 0.2 m and the model was the same as that detailed in \citep{biswas2024effect, biswas2024energy}. \rfC{The blockage based on the turbine diameter was 8.7$\%$ which is comparable to blockages encountered in previous experimental studies \citep{sherry2013interaction, miller2019horizontal}}. \rfA{The model had a nacelle (with diameter $D_{nac} = 0.033$m) and tower (with diameter $D_T = 0.021$m) associated with it such that it
resembled a utility-scale turbine}. \rfC{The total blockage (rotor + tower) was $10\%$}. \rfC{Note that in multi-megawatt turbines, the nacelle diameter is often $<5\%$ of the rotor diameter and the tower diameter is $\approx 3\%$ of the rotor diameter \citep{desmond2016description}. The design constraints in our experimental model, however, resulted in a nacelle diameter to rotor diameter ratio of $16.5\%$ and tower diameter to rotor diameter ratio of $10.5\%$. Indeed in small-scale models used in laboratory experiments it has often not been possible to have a nacelle and tower with shape and size exactly similar to full-scale turbines \citep{chamorro2013interaction, howard2015statistics, barlas2016roughness}. The turbine was vertically hung into the flume in an inverted fashion as shown in figure \ref{fig:Schematic}. A stepper motor RS 829-3512 was used along with a drive and signal generator to rotate the turbine at a prescribed RPM. The motor, along with the speed controlling electronics, were located outside the flume and the torque from the motor was transmitted to the turbine shaft via a belt and pulley mechanism. The pulley and the shaft were placed inside the nacelle and the belt ran through the hollow tower which restricted any further reduction in the diameter of the nacelle or tower.} The free-stream velocity ($U_{\infty}$) was kept constant at 0.2 ms$^{-1}$. Experiments were performed at a single tip speed ratio $\lambda_{\infty} = 6$ (defined as $\lambda = \Omega D/2U_{\infty}$, where $\Omega$ is the turbine's rotational speed). The Reynolds number based on the turbine diameter, $Re_D$ was $\approx 40000$. 

The PIV experiments with the wind turbine had the same field of view as that in experiment 1A discussed in \citep{biswas2024energy}. Therefore, the details of the experiments are only briefly discussed here. Only planar PIV experiments were performed at the diametrical plane of the rotor aligned with the tower's axis. \rfB{Three phantom v641 cameras were used to yield a stitched field of view (FOV) spanning $\approx 4.5D$ in the streamwise ($x$) direction and $\approx 1.1D$ in the spanwise ($y$) direction. The origin of the co-ordinate system was placed at the rotor plane at the hub height.} The experiments were time resolved. The acquisition frequency ($100$ Hz) was much higher than the turbine's rotational frequency ($\approx 1.87$ Hz). The spatial resolution of the experiments ($\approx 0.0115D$) was good enough to capture the tip vortices (having size $\approx 0.07D$). \rfD{Data was obtained for a total time of 54.56 seconds (5457 snapshots) which corresponded to $\approx 102$ rotor revolutions}.

 \begin{figure}
  \centerline{
  \includegraphics[clip = true, trim = 0 0 0 0 ,width= \textwidth]{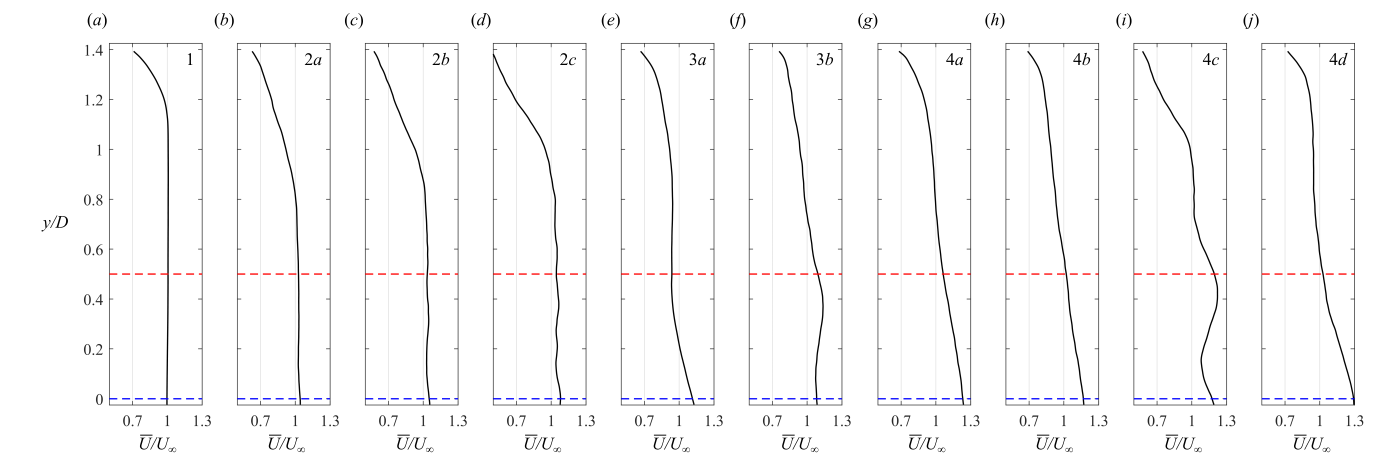}  }
 \caption{Profiles of mean streamwise velocity at the rotor plane (with the turbine removed). The case IDs are shown on the top right.  }
\label{fig:U_grid}
\end{figure}

\begin{figure}
  \centerline{
  \includegraphics[clip = true, trim = 0 0 0 0 ,width= \textwidth]{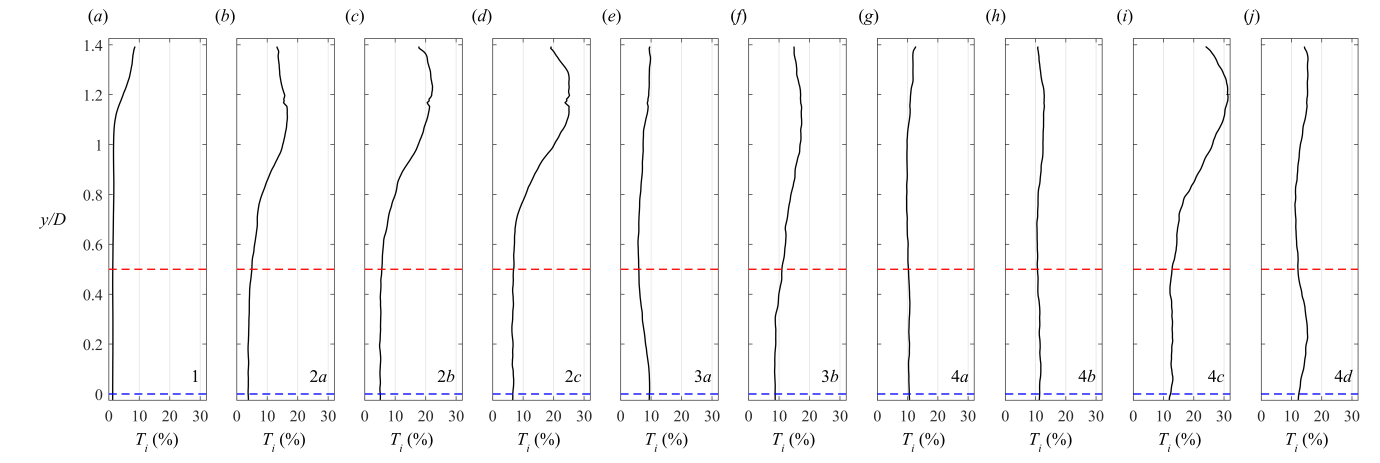}  }
 \caption{Turbulent intensity profiles at the rotor plane (with the turbine removed). The case IDs are shown on the bottom right. }
\label{fig:T_grid}
\end{figure}

\begin{figure}
  \centerline{
  \includegraphics[clip = true, trim = 0 0 0 0 ,width= \textwidth]{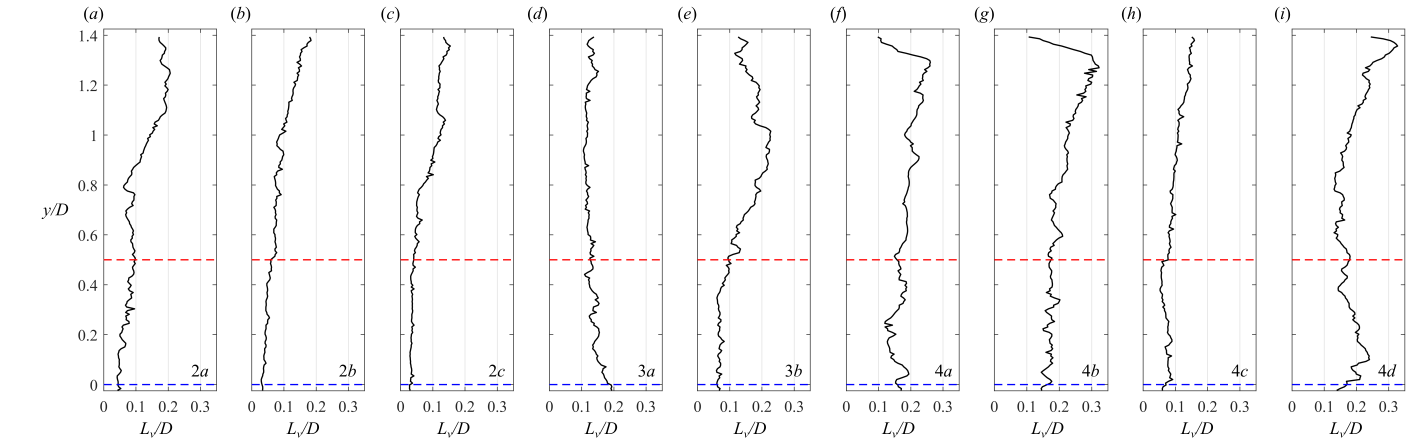}  }
 \caption{Profiles of integral length scales at the rotor plane (with the turbine removed). The case IDs are shown on the bottom right. }
\label{fig:L_grid}
\end{figure}

\rfC{In previous experimental studies, different `flavours' of freestream turbulence, primarily characterised by different turbulence intensities ($T_i$) and freestream integral length scales ($L_v$), has been produced by either using a passive turbulence generating grid \citep{blackmore2016effects, de2025influence} or using an active turbulence generating grid \citep{gambuzza2023influence, bourhis2025impact}. Active grids have been shown to be capable of producing turbulent Reynolds numbers and integral length scales much larger than those possible using passive grids \citep{kurian2009grid, hearst2015effect, hearst2018robust}. Despite limitations, passive grids have been extensively used in the literature primarily due to the ease of manufacturing and installation \citep{bearman1983effect, fransson2005transition, kankanwadi2020turbulent}. Using both types of grids however, it is generally difficult to fill the top-left and bottom-right corners of a $T_i$ - $L_v$ map, \textit{i.e.} FST with very large integral length scales and very small turbulent intensities and vice versa.}

 \begin{figure}
  \centerline{
  \includegraphics[clip = true, trim = 0 0 0 0 ,width= \textwidth]{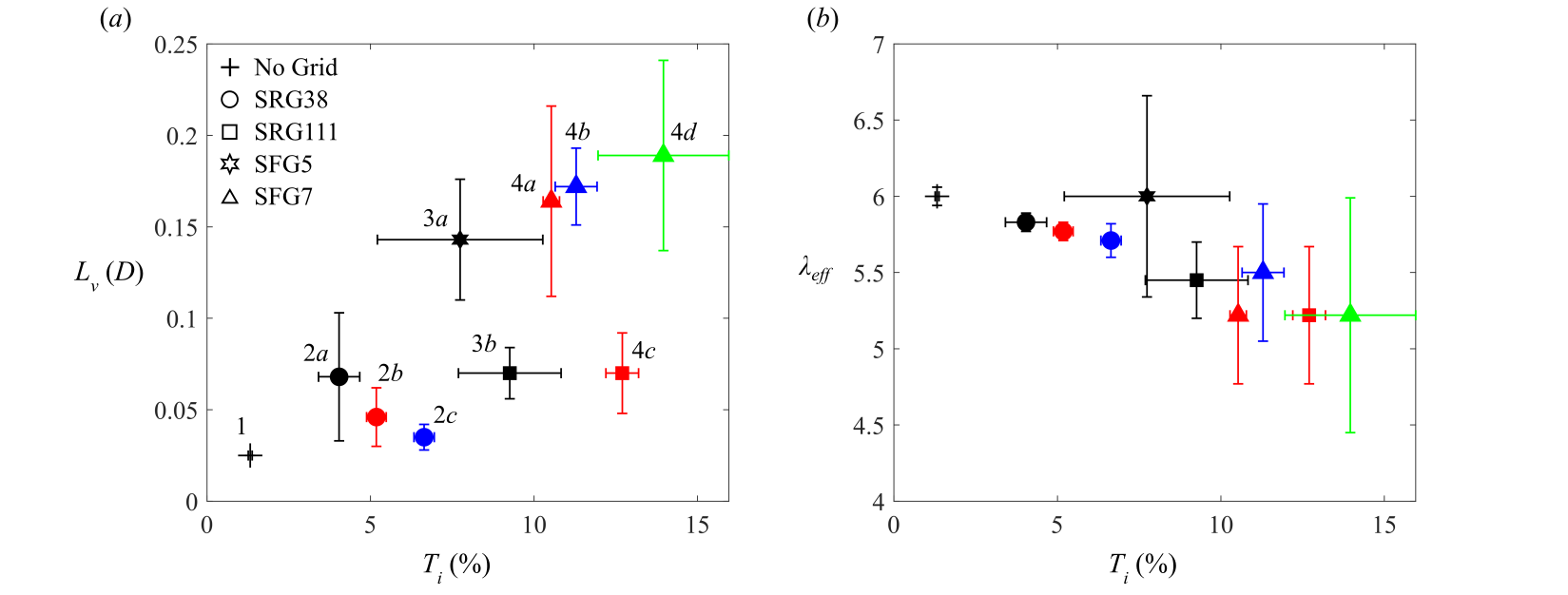}  }
 \caption{\rfC{(a) Parameter space of turbulence intensity ($T_i$) and integral length scale ($L_v$) produced by different turbulence generating grids. The symbols show the values averaged over the rotor radius. (b) shows the effective tip speed ratios ($\lambda_{eff}$) for $\lambda_{\infty} = 6$. $\lambda_{eff}$ is measured based on the rotor averaged bulk velocity, $U_b$ instead of $U_{\infty}$ and hence is lower than $\lambda_{\infty}$. The same symbol is used for a particular type of grid used as indicated in (a). The case IDs are also indicated in (a). } }
\label{fig:grid_points}
\end{figure}

\begin{table}
\centerline{
  
  \begin{tabular}{ccccccccc}
  
      Case ID & Grid   & $X_{G}$ &${T_i}$ ($\%$) & ${L_v}$ ($D$) & ${U}_b/U_{\infty}$ & $\lambda_{eff}$ & \\ \hline 
      1   & -      & -    & $1.32\pm0.06$  & $\leq 0.025$ & $1.0\pm0.01$ & 6.0 \\
      
      2a   & SRG38 & $5D$ (26.6$M$) & $4.04\pm 0.63$   & $0.068\pm0.035$ & $1.03\pm 0.01$ & $5.82\pm 0.04$ \\ 

     2b   & SRG38 & $3.5D$ (18.6$M$) &$5.18\pm0.30$   & $0.046\pm0.016$ & $1.04\pm 0.01$ & $5.78\pm 0.07$ \\ 
      
    2c   & SRG38 & $2.5D$ (13.3$M$) &$6.64\pm0.31$   & $0.035\pm0.007$ & $1.05\pm0.02$ & $5.68\pm 0.11$ \\ 

    3a  & SFG5 & $15D$ (0.47$x^*$) &$7.74\pm2.53$  & $0.143\pm0.033$ & $1.0 \pm 0.11$ & $6.01\pm0.68$  \\ 

      3b  & SRG111 & $6D$ (10.8$M$) &$9.26\pm1.57$  & $0.070\pm0.014$ & $1.10\pm0.05$ & $5.43\pm0.22$ \\ 
      
 4a  & SFG7 & $17D$ (0.58$x^*$) & $10.53\pm0.25$  & $0.164\pm0.052$ & $1.15\pm 0.10$ & $5.2\pm0.46$ \\ 

       4b  & SFG7 & $15D$ (0.51$x^*$) &$11.29\pm0.64$  & $0.172 \pm 0.021$ & $1.09\pm 0.09$ & $5.51\pm 0.46$ \\ 

       4c   & SRG111 & $3.5D$ (6.3$M$) &$12.70\pm0.50$   & $0.070\pm0.022$ & $1.15\pm 0.10$ & $5.2\pm0.45$ \\

       4d  & SFG7 & $10D$ (0.34$x^*$) &$13.96\pm2.0$  & $0.189 \pm 0.052$ & $1.15\pm0.17$ & $5.21\pm0.77$\\ 

      \hline

  \end{tabular}}
  
  \caption{Parameters associated with the different experiments. Here $X_G$ denotes the distance between the grid and the turbine. $x^*$ is the wake intersection length scale for the fractal grids \citep{mazellier2010turbulence} and $M$ is the mesh size of the regular grids. $U_b$ denotes the rotor averaged bulk velocity. $\lambda_{eff}$ is the tip speed ratio based on $U_b$.  }
  \label{tab:fst}
\end{table}

\rfC{In this work, we used four different passive turbulence generating grids to produce 9 different FST cases \rfD{(along with a reference case without a grid)} similarly to our previous studies \citep{kankanwadi2020turbulent, chen2023spatial}. Both regular and fractal grids were used (please see the supplementary material for details). Two separate experimental campaigns were carried out to obtain the variation of $T_i$, $L_v$ and the mean streamwise velocity ($\bar{U}$) behind the grids. The first experimental campaign used a single phantom v641 camera and had a thin strip like FOV of streamwise width $\approx 0.18D$, centred at a distance $X_G$ from the grid, where the turbine would be placed (see figure \ref{fig:Schematic}). The values of $X_G$ are tabulated in table \ref{tab:fst}. The transverse width of the FOV was $\approx 1.45D$, starting from close to the hub height of the turbine and reaching near the bottom wall of the flume. The profiles of $T_i$, $L_v$ and $\bar{U}$ for all the cases studied (at a distance $X_G$ from the grids) are shown in figures \ref{fig:U_grid}, \ref{fig:T_grid} and \ref{fig:L_grid} respectively. These experiments were also conducted at streamwise locations $X_G + 0.5D$ and $X_G+4D$ in order to compare with data at the same locations in the presence of the turbine. In figures \ref{fig:U_grid}, \ref{fig:T_grid} and \ref{fig:L_grid}, the blue and red dashed lines show the hub and tip heights respectively. The profiles were averaged between these two lines and the rotor-averaged quantities are used to quantify the inflow. The parametric space (spanned by $T_{i}$ and $L_{v}$) in this study is shown in figure \ref{fig:grid_points}(a) and the values are also tabulated in table \ref{tab:fst}. In figure \ref{fig:grid_points}(a), specific symbols are used for certain types of grids. The error bars show the 95$\%$ confidence interval (2 standard deviations) within the rotor radius (see the transverse variations of $T_i$ and $L_v$ in figures \ref{fig:T_grid} and \ref{fig:L_grid}) and hence mainly indicate the level of inhomogeneity in the inflow. The second experimental campaign simultaneously used 2 Phantom v641 cameras to give a stitched field of view length $\approx 2.45D$ ($49 $cm) and width $\approx 0.9D$ (18 cm). The grid was moved upstream keeping the optical setup intact and a number of non-synchronous experiments were conducted for a particular grid to obtain time averaged statistics over a certain streamwise distance as required. The results from the second campaign are only included in the supplementary material.}

\rfC{In table \ref{tab:fst}, the distance between the grid and the turbine ($X_G$) is also shown in terms mesh size ($M$) for the regular grids and the wake intersection length scale ($x^*$) for the fractal grids. Here $x^*$ is defined as $x^* = L_0^2/t_0$, where $L_0$ and $t_0$ represent the length and thickness of the largest bars of the fractal grids \citep{mazellier2010turbulence, kankanwadi2020turbulent}. For the regular grids there is a monotonic decay of turbulence and the turbulence becomes more homogeneous with downstream distance from the grid. For the fractal grids on the other hand, turbulence develops much slowly. For the latter, $T_i$ generally peaks around $x=0.45x^*$ and the flow starts to become homogeneous at around $x = 0.5x^*$ \citep{mazellier2010turbulence}. Considering several additional experimental constraints however, the grids had to be placed at locations with respect to the turbine where the inflow turbulence is not homogeneous, especially for the highest FST cases.} \rfD{ Additionally for some of the cases, especially cases 2c and 4c (closer to the regular grids) $T_i$, $L_v$ and $\bar{U}$ also had a considerable streamwise gradient. The streamwise variation of $T_i$, $L_v$ and $\bar{U}$ behind all the turbulence generating grids is shown in the supplementary material. Nevertheless, the inhomogeneity of the inflow turbulence likely only has a secondary role on the dynamics of the wake as we will be discussing. }

\rfC{One of the other factors that had to be considered while positioning the grid was the interaction of the grid with the boundary layer developing at the bottom surface of the flume. For instance, the regular grid SRG38 was found to strongly interact with the boundary layer. This was most likely because this grid had horizontal bars closest to the bottom wall of the flume (highest local blockage) in the central laser plane (see figure 1 in the supplementary material). It can be seen from figure \ref{fig:U_grid}, how the boundary layer gets progressively thicker as we move downstream of the grid (from figure \ref{fig:U_grid}(d) - \ref{fig:U_grid}(b)). Therefore, further away from the grid, although we could gain in terms of homogeneity, the boundary layer would introduce unwanted additional turbulence. For the case 2a, although the turbine was placed outside the boundary layer at the rotor plane, further downstream ($x>3D$), the turbine wake intersected the boundary layer. Therefore, for this particular case, the results were only discussed up to $x/D = 3$. This was less of an issue for the other grids. The regular grid SRG111 was intentionally placed closer to the turbine to create the cases 3b and 4c with relatively higher $T_i$ but at the expense of homogeneity. For the fractal grids on the other hand, since the turbulence develops much more slowly, a much higher $X_G$ was necessary. Now, the turbine had to be placed towards the upstream end of the test section in order to have a large portion of the wake captured in the test section measurement area (see figure \ref{fig:Schematic}). This resulted in the distance between the turbine and the upstream end of the flume being limited to $\approx$ 3m ($15D$). So the highest $X_G$ tested without shortening the streamwise extent of the wake measured was $15D$. Only for experiment 4a, the fractal grid SFG7 was placed at $X_G = 17D$, but with the wake measured up to $x/D = 3$ to see if there is a significant change in the near wake dynamics. }

 \begin{figure}
  \centerline{
  \includegraphics[clip = true, trim = 0 0 0 0 ,width= \textwidth]{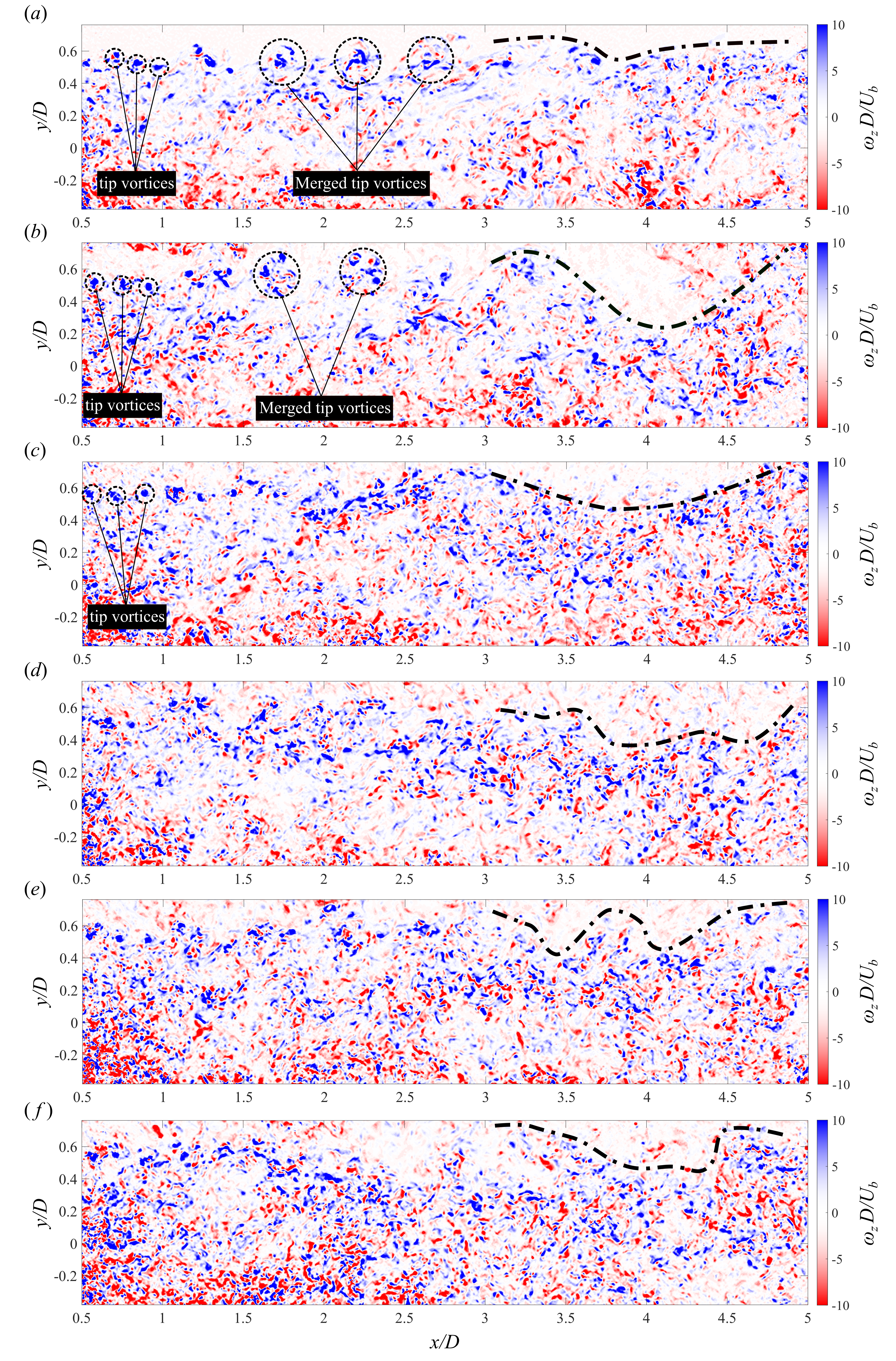}  }
 \caption{{Exemplar instantaneous} vorticity fields for the cases (a) 1, (b) 2a, (c) 2c, (d) 3b, (e) 4b and (f) 4d. The black dash-dotted line highlights the wake edge towards the far wake. }
\label{fig:42a}
\end{figure}

Estimating integral length scales involves inherent intricacies and different methods yield nonidentical results in an inhomogeneous anisotropic turbulent flow as in the present case. For instance, the integral length scales can be obtained by the direct computation of the spatial auto-correlation function or by obtaining integral time scales using the temporal auto-correlation and then invoking Taylor's hypothesis of frozen turbulence. In this work, we follow the former approach. \rfB{The integral length scales are estimated as $L_v = \int_0^{\hat{r}}R_{12}^{'}dr$, where $R_{12}^{'}$ is the transverse autocorrelation function of the streamwise velocity fluctuations. $R_{12}^{'}$ is obtained using the following equation:}

\begin{equation}
    R_{12}^{'} (\textbf{x},\textbf{y},r) = \frac{\overline{u_1'(\textbf{x}, \textbf{y}) u_1'(\textbf{x}, \textbf{y}+r \textbf{e}_2)}} {\sqrt{\overline{u_1'(\textbf{x}, \textbf{y})^{2}}} \sqrt{\overline{u_1'(\textbf{x}, \textbf{y}+r \textbf{e}_2)^{2}}}} 
\end{equation}

\rfB{Here the overbars denote averaging over time. $u_1'$ is the streamwise velocity fluctuation and $r \textbf{e}_2$ represents some displacement $r$ along the $\textbf{e}_2$ direction ($\textbf{e}_2$ is the unit vector along the spanwise, $y$ direction). Finally $L_v$ is calculated by integrating $ R_{12}^{'}$ up to a certain distance $\hat{r}$ where $ R_{12}^{'}$ first becomes zero. Further details about the quantification of the integral length scales and the sources of uncertainties have been added in appendix 1 and in the supplementary material.} It is worthwhile mentioning that other transverse or streamwise autocorrelation functions were also tested. It was seen that the streamwise autocorrelation function of the transverse velocity ($R_{21}^{'}$) yielded integral length scales of magnitude similar to that obtained using {$R_{12}^{'}$ whilst,} the streamwise autocorrelation function of the streamwise velocity ($R_{11}^{'}$) yielded larger (up to $2-3$ times) integral length scales. However the autocorrelation functions $R_{12}^{'}$ or $R_{21}^{'}$ more often fell off to zero compared to $R_{11}^{'}$ {which exhibited long range correlations}. Therefore, only the integral length scales calculated using $R_{12}^{'}$ are discussed here. A more detailed discussion on the different correlations can be found in the supplementary material. \rfB{Nevertheless, the largest integral length scales we are able to produce here ($<0.2D$ using $R_{12}^{'}$ or $<0.45D$ using $R_{11}^{'}$) are an order of magnitude smaller than the largest length scales a real wind turbine residing in an atmospheric boundary layer can be exposed to \citep{larsen2007dynamic, porte2020wind}. {The current study is thus only able to} explore the effect of short FST length scales on the near wake dynamics of a turbine.}

The grids also introduce a blockage to the freestream, resulting in an acceleration of the incoming flow \rfC{(please see the supplementary material)}. Although the tip speed ratio based on $U_{\infty}$ (denoted as $\lambda_{\infty}$) was kept constant at 6 for all the FST cases, due to the blockage induced increment of the freestream velocity (see figure \ref{fig:U_grid}), the effective tip speed ratio (defined as $\lambda_{eff} = \lambda_{\infty} U_{\infty}/U_b$) reduced for all the FST cases. Here $U_b$ is the rotor averaged bulk velocity. $\lambda_{eff}$ for the different cases is shown in figure \ref{fig:grid_points}(b). The uncertainties are due to the transverse variation of the mean streamwise velocity as shown in figure \ref{fig:U_grid}. \rfE{The variation of the power ($C_P$) and thrust coefficients ($C_T$) as a function of the tip speed ratio were obtained in our previous work using the blade element momentum method \citep{biswas2024effect} (please see figure 15(b)). For the range of $\lambda_{eff}$ observed here, the variation in the $C_P$ and $C_T$ was expected to be marginal.} The test cases are divided into 4 subgroups, primarily based on turbulence intensity. Group 1 has negligible $T_i$ while groups 2, 3 and 4 have low, mild and high $T_i$ respectively. Nevertheless, hard boundaries {likely do not }exist between the adjacent subgroups.

\section{Effect of freestream turbulence on the dynamics of the wake}
\label{near}

Instantaneous vorticity fields for 6 cases, \rfD{case 1 (no grid case)}, 2a, 2c, 3b, 4b and 4d are shown in figure \ref{fig:42a} for $\lambda_{\infty}=6$ (see also supplementary videos 1-6). These cases represent the wake at increasing turbulence intensities. \rfE{The instantaneous vorticity fields show a rich variety of scales present in the flow.} \rfB{The individual tip vortices and the merged vortices are the most clearly observed for the no grid case (case 1) as highlighted in figure \ref{fig:42a}(a). Here, the tip vortices undergo merging at $x/D \approx 1.5$ similarly to {the observation} in our earlier study through a different set of experiments \citep{biswas2024effect}. As $T_i$ is increased, the individual tip vortices can still be observed (see figure \ref{fig:42a}(b,c)), but it becomes increasingly difficult to detect the merging process and the merged tip vortices. In general, as $T_i$ is increased, the tip vortices break down much closer to the turbine as reported in previous studies \citep{chamorro2009wind, wu2012atmospheric, gambuzza2023influence}. For the no grid case, the wake edge towards the far wake (indicated by the black dash-dotted line) stays nearly flat. However, for the other cases the wake edge shows significantly larger undulation. This can be seen more clearly from the supplementary videos 1-6. We look at the important frequencies present at different locations in the wake to better understand these primitive observations. }

\begin{figure}
  \centerline{
  \includegraphics[clip = true, trim = 0 0 0 0 ,width= \textwidth]{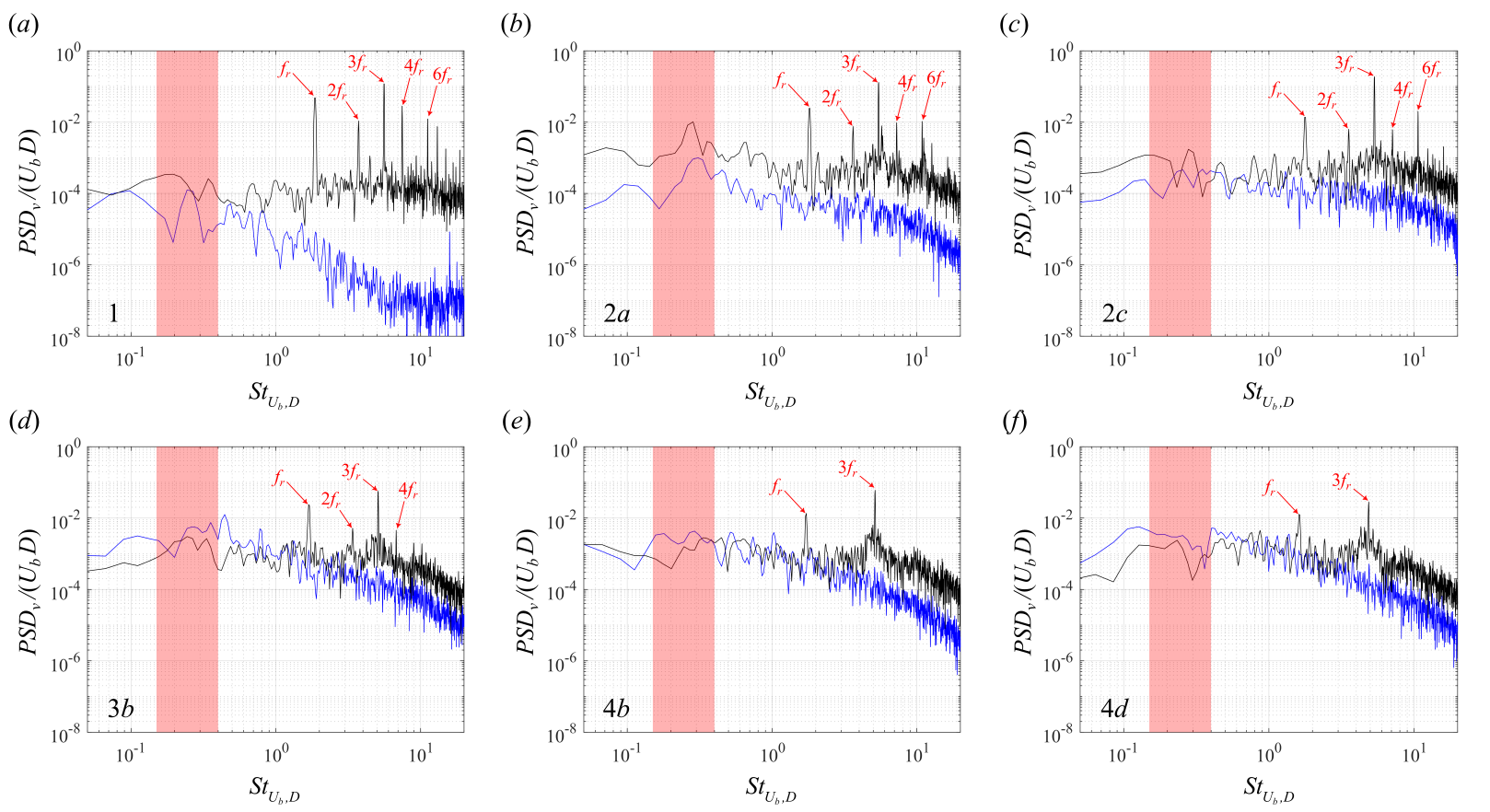}  }
 \caption{\rfC{Spectra at $x/D = 0.5$ and $y/D = 0.55$ for the cases (a) 1, (b) 2a, (c) 2c, and (d) 3b, (e) 4b and (f) 4d. The spectra with and without the turbine are shown in black and blue respectively. The shaded region shows the wake meandering frequency band ($0.15\leq St_{U_{b},D}\leq 0.4$).} }
\label{fig:43a}
\end{figure}

\begin{figure}
  \centerline{
  \includegraphics[clip = true, trim = 0 0 0 0 ,width= \textwidth]{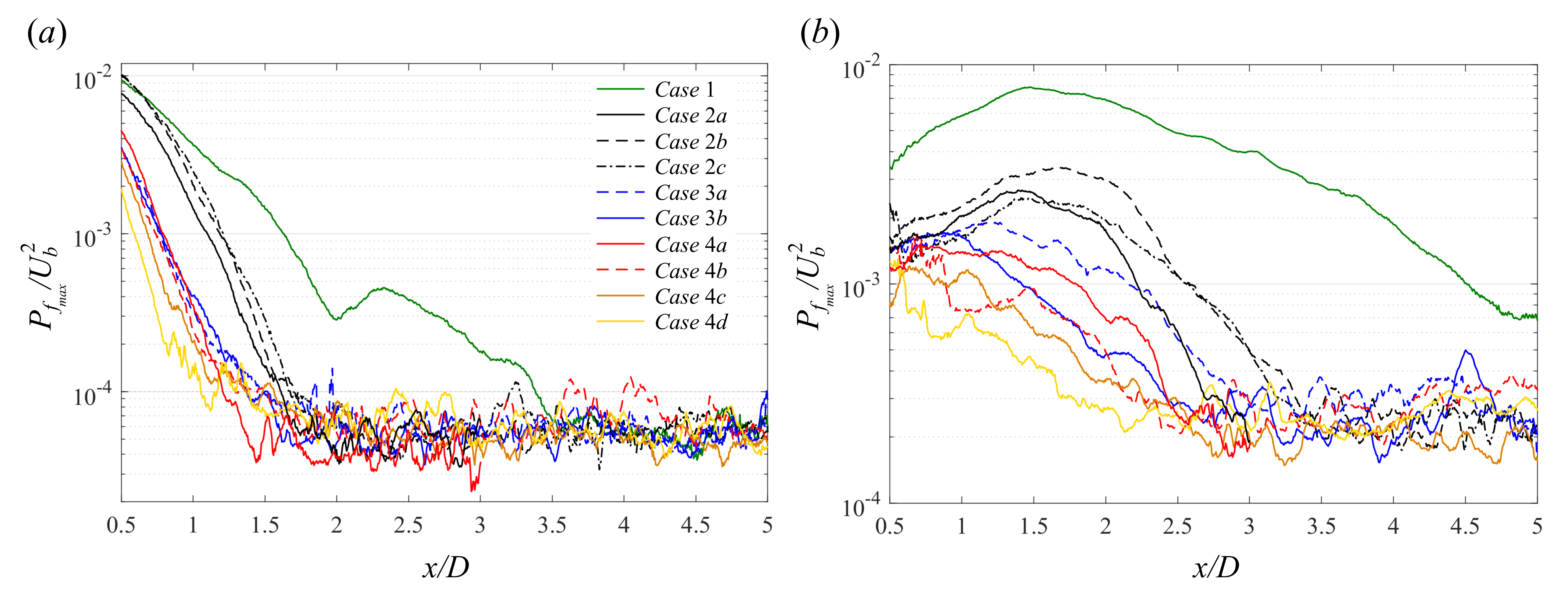}  }
 \caption{\rfC{Maximum power of the tip vortex related frequencies (a) $3f_r$ and (b) $f_r$ for the different FST cases.}}
\label{fig:tip_Sf}
\end{figure}

\subsection{Important frequencies}


 \rfB{In order to understand the effect of FST on the {coherent motions at different frequencies}, {we look at the PSDs of transverse fluctuating velocity ($PSD_v$).} \rfD{This is first shown in the tip region ($y\approx 0.5D$) close to the rotor ($x/D=0.5$, $y/D=0.55$)} in figure \ref{fig:43a} for the different cases (in black).} \rfC{Here the PSDs are estimated using Welch's method with a window size of $N/2$ (where $N=5456$ was the length of the time series) \citep{welch1967use}. \rfB{This yielded a frequency resolution of 0.0244 hz ($St_{U_{\infty},D} = 0.0244$).} The spectra at the same location without the turbine is also shown in blue for comparison. The Strouhal number, $St_{U_b, D}$ is calculated based on the rotor averaged bulk velocity ($U_b$) and the turbine diameter ($D$).} \rfB{For all the FST cases, the dominant frequency is $3f_r$, where $f_r$ is the turbine's rotational frequency ($\approx$1.87 hz).} $3f_r$ therefore denotes the frequency of the tip vortices for a three-bladed turbine. Apart from $3f_r$, $f_r$ is also found to be important and is observed in all cases. The harmonics of $f_r$ up to $6f_r$ and even beyond are clearly observed for case 1. \rfD{The red band in the low frequency region $0.15\leq St_{U_b,D} \leq 0.4$ shows the typical frequency range of wake meandering reported in previous studies \citep{medici2008measurements, okulov2014regular, chamorro2013interaction}}. Note that for all the cases with a turbulence generating grid, the spectral energy content in this wake meandering frequency band (WMFB) is higher than the \rfD{no grid case}. More interesting observations can be made by comparing the spectra at the same location without the turbine. For the cases with low FST, the turbine seems to add energy in the WMFB (figures \ref{fig:43a}(a-c)). Whilst, for medium to high FST (figures \ref{fig:43a}(d-f)), there is an attenuation of energy in the wake in this low frequency region. Especially for case 2a, there is a clear peak in the vicinity of $St_{U_b,D}$ that is present both in the freestream and in the wake and is amplified in the wake. This clearly shows how the nature of the inflow turbulence can impact the low frequency wake meandering dynamics in the wake. This will be discussed in more detail in the next sections. 
 

For the low $T_i$ cases 2a and 2c, most of the harmonics of $f_r$ are still observed in the tip region similar to case 1.  For the medium $T_i$ case 3b, the harmonics such as $2f_r$ and $4f_r$ are observed, although their amplitudes are significantly diminished. $6f_r$ is no longer observed clearly for case 3b. Finally, for the highest $T_i$ cases 4b and 4d, {no spectral peaks are evident other than at $f_r$ and $3f_r$ within the field of view considered here {($0.5 \leq x/D \leq 5$)}}. The other frequencies can however still be present very close to the turbine blades.

 \begin{figure}
  \centerline{
  \includegraphics[clip = true, trim = 0 0 0 0 ,width= \textwidth]{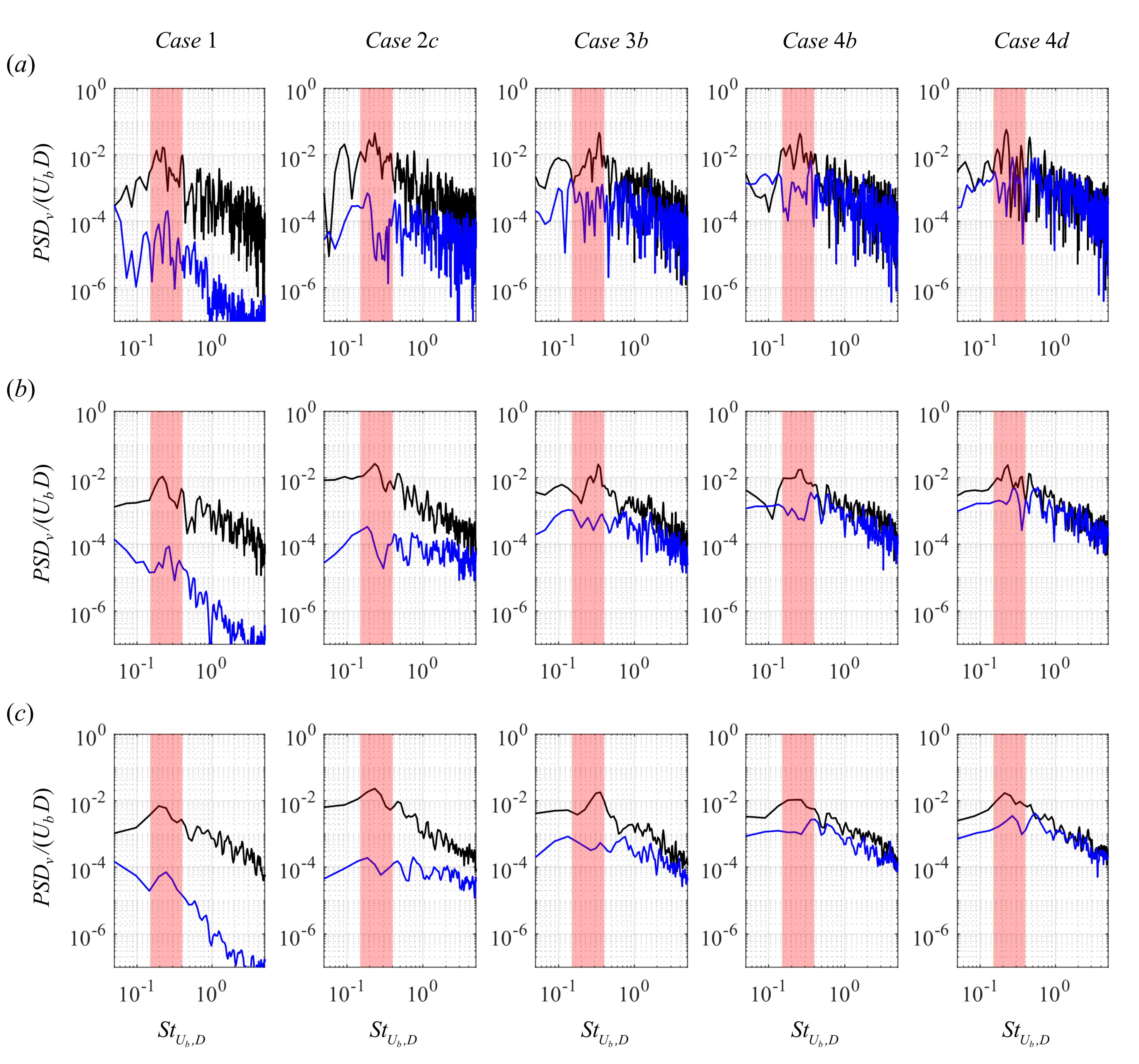}  }
 \caption{\rfC{Spectra at $x/D = 4$ and $y/D = 0$ obtained using Welch's PSD estimator having window lengths (a) $N$, (b) $N/2$ and (c) $N/4$ for the cases 1, 2c, 3b, 4b and 4d (stacked column wise). Here $N$ corresponds to the total length of the time series. The windows had 50\% overlap. The spectra with and without the turbine are shown in black and blue respectively. The shaded region shows the wake meandering frequency band ($0.15\leq St_{U_{b},D}\leq 0.4$).}}
\label{fig:pwelch_fwm}
\end{figure}

\rfC{Figure \ref{fig:43a} showed that $f_r$ and $3f_r$ are the only two tip vortex related frequencies that could be observed for all the FST cases. Therefore, we explore how the amplitude of these two frequencies vary in the streamwise direction for the different cases. The amplitude was measured in terms of the power contained in a frequency peak ($P_f(x,y)$) at a particular location. The power was calculated by integrating the PSD over a $\Delta f =0.14$hz $\Delta St_{U_{\infty},D} =0.14$) which was the best estimate of the width of the spectral peaks associated with tip vortex related frequencies ($f_r$, $3f_r$ etc). The maximum value of $P_f(x,y)$ at a particular $x$ was taken and the streamwise variation of the maximum power ($P_{f_{max}}$) is shown in figure \ref{fig:tip_Sf}. As can be expected, $3f_r$ or the tip vortices sustain the furthest downstream for the no grid case. As the FST level is increased they decay faster and closer to the turbine. Interestingly, among the low $T_i$ cases (group 2), $3f_r$ decays the fastest for case 2a which had the lowest $T_i$ and the highest $L_v$ in group 2. In fact, for the majority of the cases, the decay of $3f_r$ can be seen to be more correlated to $L_v$ than $T_i$. The formation of $f_r$ is more complex and energy can be introduced at this frequency due to for instance the blade eccentricities (very close to the turbine) and due to the merging of the tip vortices further downstream \citep{abraham2023experimental, biswas2024effect}. A preferred dependence on either $T_i$ or $L_v$ could not observed clearly for $f_r$. Nevertheless, it decayed closer to the turbine with increasing FST levels. }

\rfC{Figures \ref{fig:pwelch_fwm} shows the PSDs at $x/D = 4$ and $y/D=0$ for the five cases (excluding case 2a) of increasing turbulence intensity. At this location, the wake meandering frequency can be expected to become important. Note that the tip vortex related frequencies always showed discrete peaks and therefore their detection was largely insensitive to how the spectrum was calculated. In contrast, the detection of the low frequency peaks in the WMFB (shown by the red shaded region in figure \ref{fig:pwelch_fwm}) was more sensitive to the choice of the method and the associated parameters, due to their less discrete nature. Therefore, to better understand the nature of the spectra in the WMFB, the spectra were computed with different window sizes in the Welch's method. In figure \ref{fig:pwelch_fwm}, the rows a, b and c correspond to window lengths of $N$, $N/2$ and $N/4$ respectively (where $N=5456$ is the length of the time series). For the highest window length, the spectra were relatively noisy, whilst the smallest window length oversimplified the spectra. Therefore, all the analysis related to PSD was carried out with the in-between window length of $N/2$ with a $50\%$ overlap. Nevertheless, clear peaks are observed in the WMFB for all the cases, even with the highest FST levels. Furthermore, the amplitude of the peak in the WMFB for all the FST cases were stronger compared to the no grid case (see row (b) in figure \ref{fig:pwelch_fwm}). This is in line with the previous studies who also reported that the amplitude of wake meandering increased {in the presence of} FST \citep{espana2011spatial, gambuzza2023influence}. \textcolor{red}{The clearest peak was however observed for the no grid case at $St_{U_b,D} \approx 0.2$. For the other cases, this peak is less coherent or more broadband in nature}. This is similar to the observation of \citet{bourhis2025impact} who noted a broadening of the wake meandering frequency band for high $T_i$. Note that for the no grid case, there is another peak observed at $St_{U_b,D} \approx 0.4$. This peak was likely a signature of the decaying hub vortex. Although some previous studies have found the hub vortex to become incoherent at around this location \citep{foti2016wake}, the relatively larger nacelle used in this study likely creates a stronger hub vortex that sustains further downstream (see for instance figure 4(c,d) of \citep{biswas2024energy}). The freestream spectra calculated at the same location is also shown in blue. Note that for all the cases the turbine has added energy in the WMFB. The energy difference between the freestream and the wake reduces with increasing FST level. This indicates that in the central region of the far wake, wake meandering is still driven primarily by the turbine, not FST.  }

\begin{figure}
  \centerline{
  \includegraphics[clip = true, trim = 0 0 0 0 ,width= 0.9\textwidth]{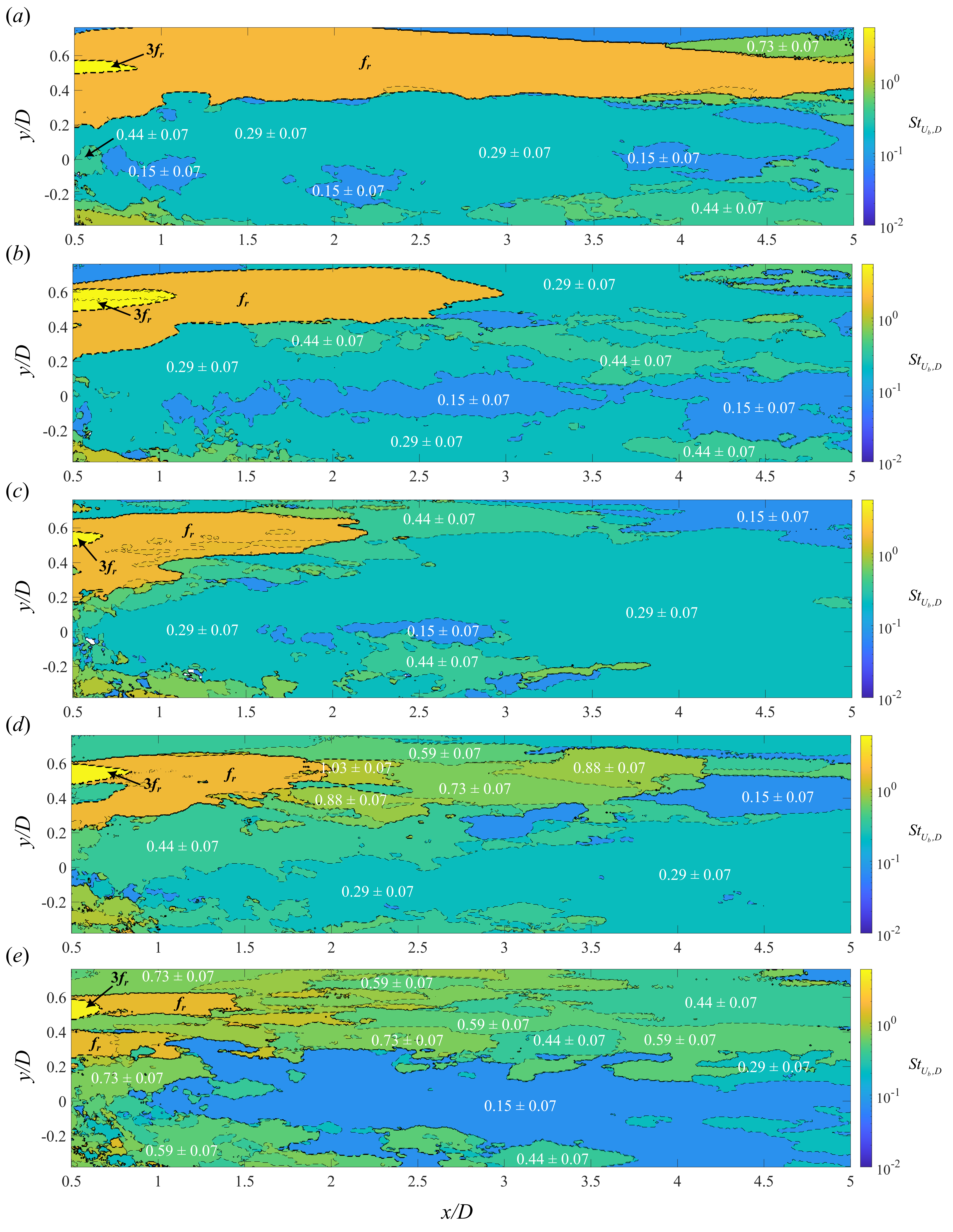}  }
 \caption{{Dominant frequency maps for the cases (a) 1, (b) 2c, (c) 3b, (d) 4b and (e) 4d obtained based on the power contained in a narrow frequency band. The power spectra are obtained using Welch's PSD estimation with a window length of $N/2$ ($N=5456$ being the length of the time series).}}
\label{fig:43c}
\end{figure}

{\subsection{Frequency maps}}

\rfC{For a global view of the important frequencies in different parts of the wake, a dominant frequency map is created {similarly to that reported in \citep{biswas2024effect}}. These are shown in figure \ref{fig:43c} for the five cases of increasing turbulent intensity. Here, we identified the dominant frequency at a spatial location based on the total power contained around that frequency. The power spectra was integrated over a frequency window $\Delta f$ centred around the frequency in question. If $\Delta f$ is too large there would be nonphysical contributions from different frequencies whilst a $\Delta f$ too small would under predict the power in a particular coherent frequency band. $\Delta f =0.14$ hz ($\Delta St_{U_{\infty},D} =0.14$) was chosen which was the best estimate of the width of the spectral peaks associated with tip vortex related frequencies ($f_r$, $3f_r$ etc). A slight variation of $\Delta f$ didn't change the final nature of the maps. }

\rfC{The look of the maps also depends on the number/values of the contours chosen. We showed specific contours with two contours separated by a minimum of $\Delta f$, \textit{i.e.} the frequencies are binned with a bin size of $\Delta f$. A frequency zone is represented just by the mid frequency in the bin with frequency boundaries $f\pm \Delta f/2$. These frequency boundaries are larger compared to the uncertainty caused by the variation in $U_b$ for all the frequencies of importance. }

 The map of the non-turbulent case for $\lambda=6$ is shown in figure \ref{fig:43c}(a) and it looks similar to that reported in \citep{biswas2024effect} {obtained through} a different set of experiments. In figure \ref{fig:43c}(a), there is a small region close to the tip near the rotor where the tip vortices ($3f_r$) are the strongest. However, they soon interact with each other producing $f_r$, which remains the dominant frequency in the {wake shear layer} even beyond $5D$. {The frequencies related to the tip vortices, especially $f_r$, remain dominant in a nearly axisymmetric shell type region apart from the region behind the tower \citep{biswas2024effect}}. 
This shell type region demarcates the inner wake from the outer background fluid. In the inner wake region, multiple low frequencies dominate in large patches and these frequencies fall in the accepted frequency range of wake meandering.

\rfD{For the cases with FST, the high frequency shell breaks down much closer and a connection between the outer turbulent flow and the inner wake is established much earlier. This has consequences in terms of the sustainment of wake meandering in the far wake for these case as we will discuss. For the highest turbulence intensity cases, many new low frequencies appear in smaller patches, making it look more broadband and turbulent-like in the inner wake region as well as at the wake's outer edge. The interaction with these frequencies likely makes wake meandering more broadband in the far wake.}  \\

\begin{figure}
  \centerline{
  \includegraphics[clip = true, trim = 0 0 0 0 ,width= \textwidth]{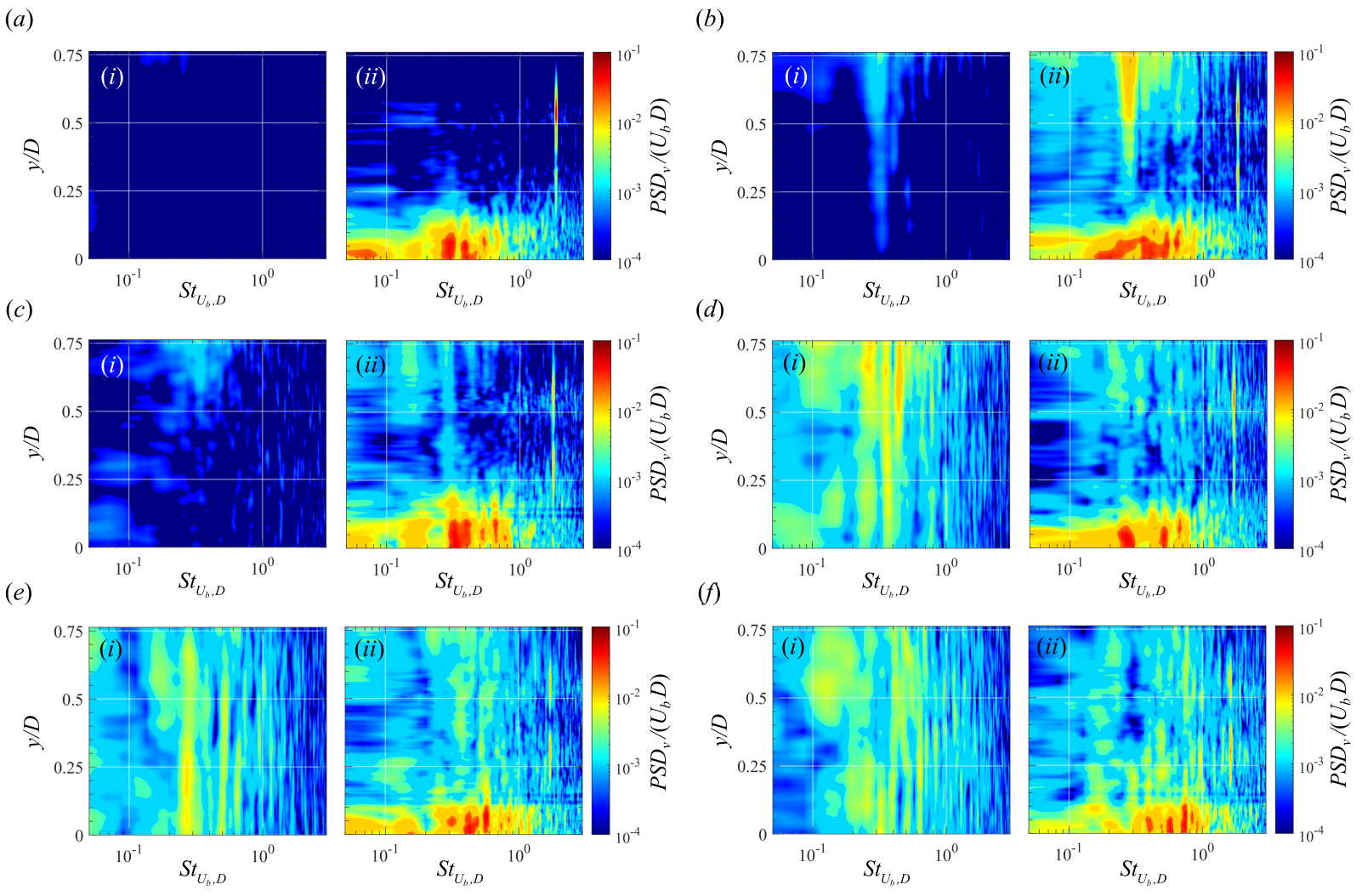}  }
 \caption{PSDs in the freestream (without the turbine) and in the wake, $0.5D$ downstream of the turbine for (a) no grid case, (b) case 2a, (c) case 2c, (d) case 3b, (e) case 4b and (f) case 4d, obtained using Welch's PSD estimation with a window length of $N/2$.  }
\label{fig:freestream_wake_fft}
\end{figure}

{\subsection{The nature and amplitude of wake meandering in the presence of FST}}

\label{wm}

To elucidate the nature of wake meandering in the presence of FST, {power} spectral densities are shown in figure \ref{fig:freestream_wake_fft} with and without the turbine. In each figure set, sub figure (ii) shows the spectrum in the turbine wake at a distance of $x/D=0.5$ from the turbine itself. Sub figure (i) shows the spectrum at the same location with the turbine removed. The spectra correspond to the same six cases (1, 2a, 2c, 3b, 4b and 4d) as discussed previously. The energy content in the freestream spectra is the lowest for the no grid case and it increases for the high $T_i$ cases as expected. \rfA{For all the cases in the turbine wake, low frequency peaks can be observed in the central region (near $y=0$) in the range $0.15\lesssim St_{U_{b},D} \lesssim 0.4$ and $0.4\lesssim St_{U_{b},D} \lesssim 0.6$. The former frequency falls in the wake meandering frequency band defined earlier. The latter matches well with the hub vortex or nacelle shedding frequency reported in several previous studies \citep{iungo2013linear, viola2014prediction, howard2015statistics, abraham2019effect}. The co-existence of the two frequencies in the vicinity of the nacelle is interesting. It is possible that the lower frequency is a subharmonic of the nacelle's vortex shedding that grows into wake meandering further downstream. This is in line with \citep{foti2016wake} who showed that the meander profiles in the near wake ($x<2D$) followed the hub vortex}. Some energy content is also observed in the frequency range $0.6\lesssim St_{U_{b},D} \lesssim 1$ which is likely associated with the altered vortex shedding from the tower \citep{de2021pod, biswas2024effect}

\rfA{Interesting observations can be made by looking at the nature of the spectra in the tip region ($y/D \approx 0.5$). Unlike the no grid case, for the other cases, there is a non-negligible energy content in the WMFB in the tip region. Interestingly, the low $T_i$ case (case 2a) showed the highest energy content in the WMFB in the tip region. The spectra without the turbine clearly shows some non-negligible energy content in almost the same frequency range that appears to have been amplified in the wake of the turbine. Curiously, there is more energy in the freestream in the WMFB for some other cases such as the 3b, 4b and 4d. However, there is a clear attenuation of the low frequencies in the wake for these cases as observed in previous studies \citep{chamorro2012evolution, jin2016effects}. This is most likely due to the fact that for these cases, the freestream spectra in this region is more multiscale in nature (presence of multiple frequencies) in contrast to more of a single frequency for case 2a. Case 2a therefore represented more of an idealised inflow where there is energy localised in a narrow band. An example of such inflow could be the flow behind a hill, where there is a coherent vortex shedding that might interact with turbines.} \rfC{Now, for case 2a, the turbine is placed $5D$ or $26.6M$ ($M$ being the mesh size) behind the regular grid SRG38, where the flow should be well developed. This frequency is therefore most likely introduced by the modulation of the outer edge of the boundary layer developing on the bottom wall of the flume, although the turbine wasn't strictly placed within the boundary layer (please see figure \ref{fig:U_grid}(b)). As was discussed earlier, the grid SRG38 strongly interacted with the boundary layer. For case 2a, the distance of the rotor from the grid was the highest ($5D$) and the height of the boundary layer was also the highest $\approx 0.63D$ for this case (see figure \ref{fig:U_grid}(b)). Therefore, this frequency, although not strictly produced by the grid alone, shows interesting physics in this case.} 

 These observations are in line with those from recent works which suggest that the turbine has a dynamic response \citep{mao2018far, heisel2018spectral, gupta2019low}, \textit{i.e.} the details of the inflow spectra is important in determining the wake dynamics. {It is worthwhile to note that FST can also indirectly induce/influence wake meandering through breaking down the tip vortices \citep{howard2015statistics}.} Assessing the relative importance of FST in promoting wake meandering directly or indirectly through the tip vortices is non trivial and deserves future investigations. Nevertheless, some evidence of interaction between the tip vortices and wake meandering in the presence of FST is presented in the next paragraphs.



 \begin{figure}
  \centerline{
  \includegraphics[clip = true, trim = 0 0 0 0 ,width= 0.9\textwidth]{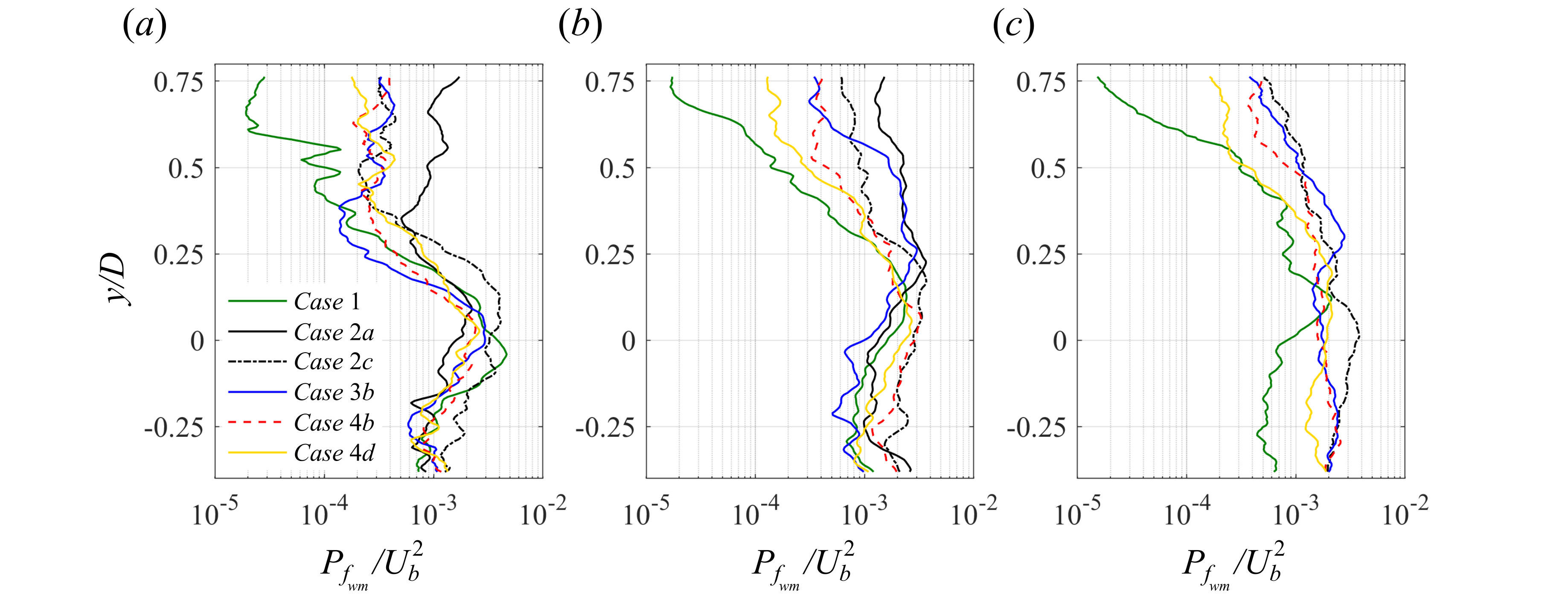}  }
 \caption{{Transverse variation of power in WMFB for different FST cases at (a) $x/D=1$, (b) $x/D=3$ and (c) $x/D=5$. }}
\label{fig:a_wm_profile}
\end{figure}

 \begin{figure}
  \centerline{
  \includegraphics[clip = true, trim = 0 0 0 0 ,width= \textwidth]{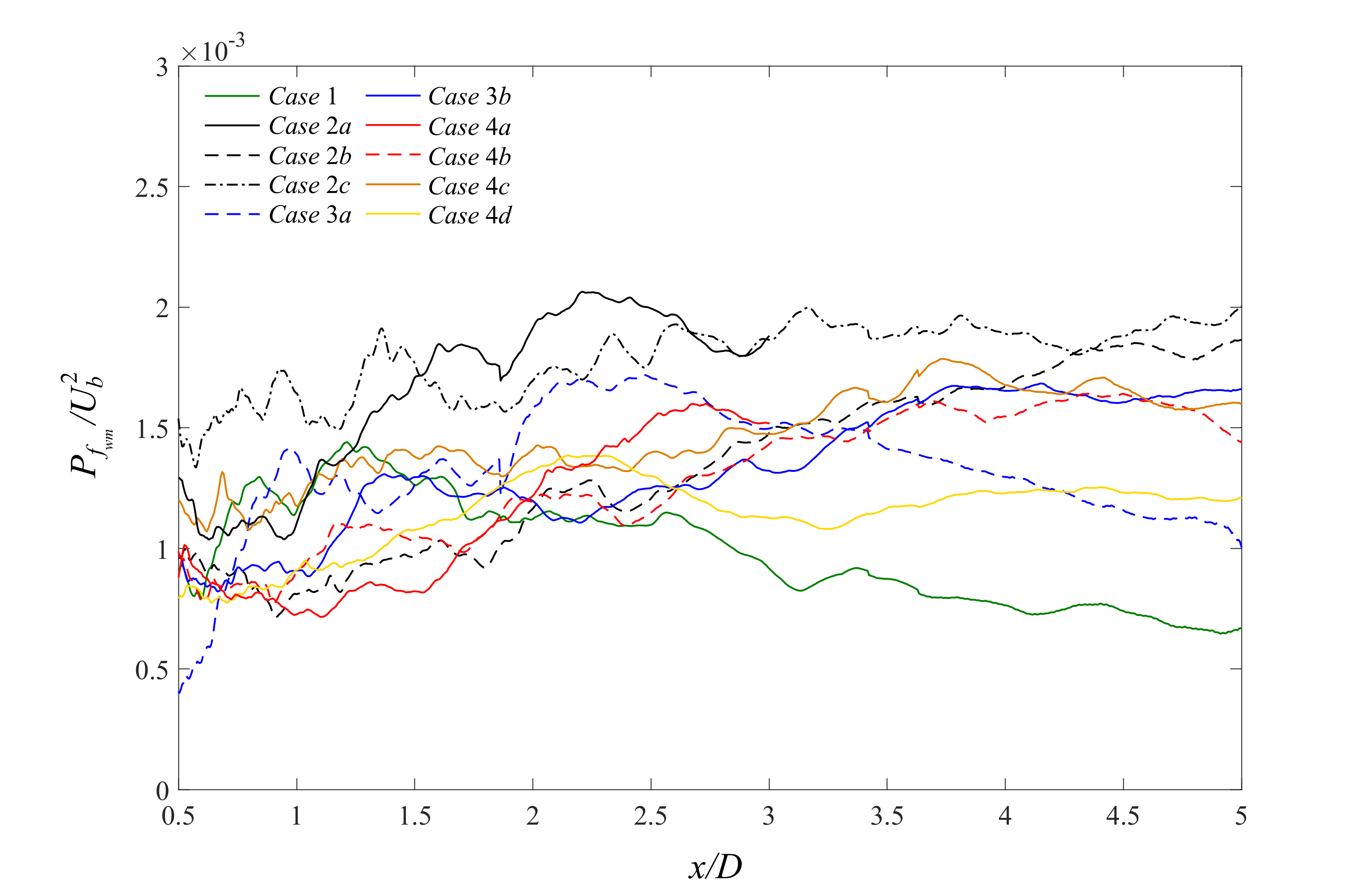}  }
 \caption{{Streamwise variation of average power (averaged in $y$ direction) in WMFB for different FST cases. } }
\label{fig:a_wm_avg}
\end{figure}

 \begin{figure}
  \centerline{
  \includegraphics[clip = true, trim = 0 0 0 0 ,width= \textwidth]{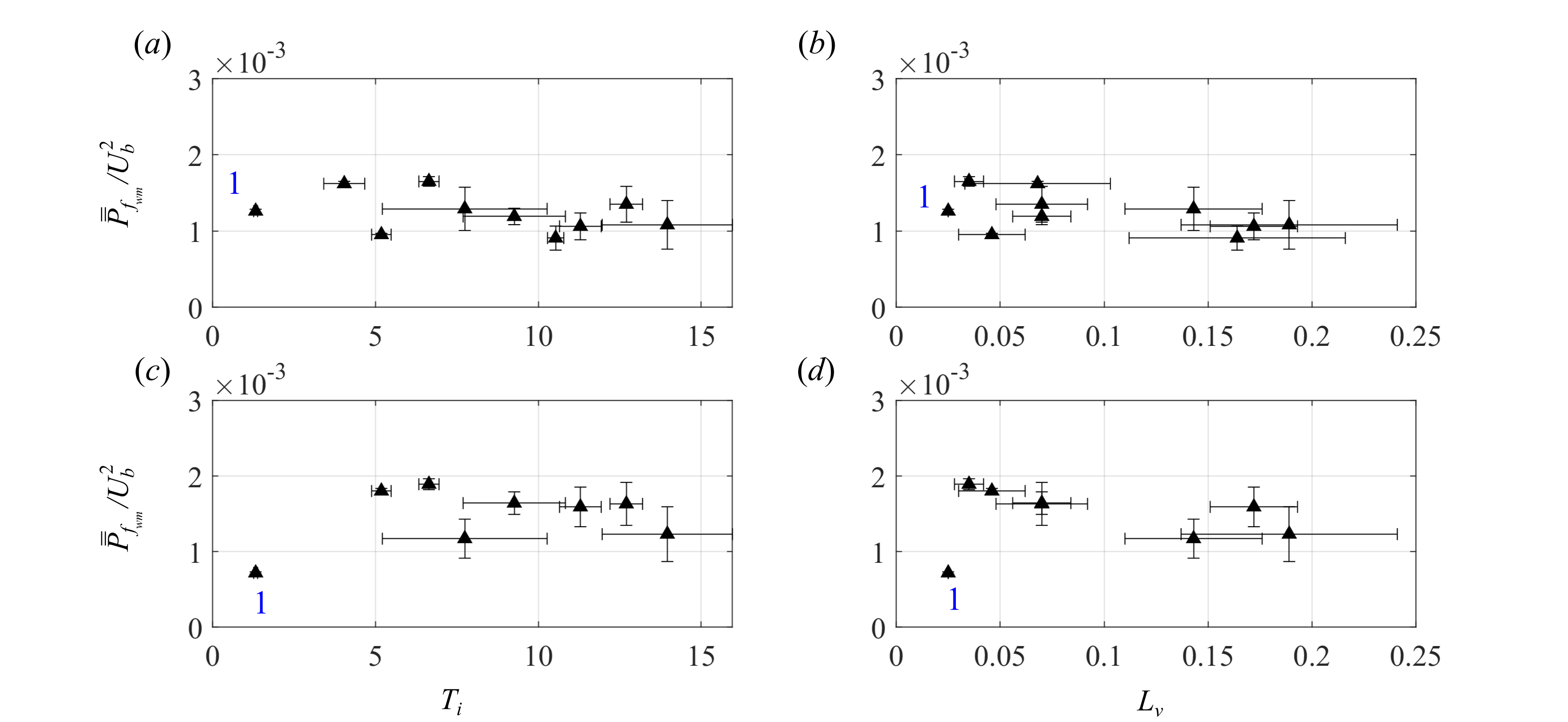}  }
 \caption{ Variation of average power in the WMFB (a, b) in the near wake (averaged between $x/D = 1$ and $x/D=2$) and (c, d) in the far wake (averaged between $x/D = 4$ and $x/D=5$) for different FST cases. }
\label{fig:a_wm_avg_avg}
\end{figure}

\rfA{Next we explore how the total power in the WMFB varied for the different FST cases. The nacelle's shedding frequency at $\lambda=6$ was previously found to be around $St_D\approx0.42$ \citep{biswas2024effect}. Therefore, in order to avoid unwanted contributions from the nacelle's shedding frequency, the power spectra was integrated over a frequency range $0.15\leq St_{U_b,D}\leq 0.37$ (containing $N_f \approx 10$ resolved spectral peaks) only. The wake meandering power profiles hence created are first shown at $x/D=1$ in figure \ref{fig:a_wm_profile}(a) for selected cases. In the central part of the wake, \textit{i.e.} in the proximity of the nacelle, the power profile for the no grid case shows the strongest peak. The amplitude thereafter decays towards the edge of the wake. This reaffirms the role of the nacelle in initiating wake meandering when the FST level is negligible. For the other cases with FST, the profiles still show a peak in the central region, although diminished, highlighting that the nacelle is still the key in inducing wake meandering in the central wake region for the FST cases considered here. Unlike the no grid case however, the FST cases show non-negligible wake meandering power towards the edge of the wake. In line with the observation in figure \ref{fig:freestream_wake_fft}(b), case 2a shows the highest power in the tip region in the near wake. For the mid to high FST cases, although figure \ref{fig:a_wm_profile} shows more power compared to the no grid case, the turbine actually attenuated the power already available in the freestream in the WMFB (as was seen in figure \ref{fig:freestream_wake_fft}). Further downstream (figures \ref{fig:a_wm_profile}b,c), the power in the central region diminishes for the no grid case, while that for the other cases is enhanced. }

\rfA{To compare all the cases simultaneously, the power profiles are averaged over $y$ and are shown for all the cases considered here in figure \ref{fig:a_wm_avg}. Note that nearer to the turbine, for $x/D \leq 2.5$, the average wake meandering powers are more comparable for all the cases. Beyond $x/D = 2.5$, the differences become more discernible. In particular, a monotonic decline is observed for the no grid case. Note that for the no grid case, the low frequency region in the inner wake remains shielded by the shell type region dominated by $f_r$ (see figure \ref{fig:43c}(a)) throughout the streamwise extent of the FOV. For this case wake meandering is likely triggered by the hub vortex/nacelle's shedding and it follows a trend similar to the decay of the hub vortex \citep{foti2016wake}. For the cases with higher FST level, the average powers of wake meandering are similar to the no grid case in the near wake. However, unlike the no grid case, further downstream, wake meandering is sustained for the other cases. This has to do with the earlier breakdown of the tip vortices as depicted in figure \ref{fig:tip_Sf}. The breakdown of the tip vortices likely provides additional impetus to sustain far field wake meandering. \citet{neunaber2020distinct} divided the wake into four distinct regions. These included the near wake where the centrline $T_i$ decayed (the nacelle's influence was felt) and immediately downstream, a transition region where the tip vortices break down and the shear layer starts to expand. The extents of these locations depended on the FST levels. Following the same terminology, we can say that for the cases with FST here, the transition region ensues much earlier. For the no grid case, the delayed break up of the tip vortices might eventually give a boost to wake meandering further downstream which is not captured in the current FOV. Whether or not wake the meandering amplitude becomes comparable further downstream is an interesting question and deserves further investigation.   }

\rfA{To summarise the results, the spanwise averaged power curves in figure \ref{fig:a_wm_avg} are further averaged over $x$ in the intervals $1<x/D<2$ (near wake) and $4<x/D<5$ (far wake) and shown in figure \ref{fig:a_wm_avg_avg}. In the near wake, the average powers do not show any clear trend with either $T_i$ or $L_v$ and are nearly the same for all the cases. This observation confirms that the primary cause of wake meandering here is the turbine itself, not the FST. Further downstream, for all the cases with FST, the average power of wake meandering is discernibly higher than that of the no grid case. This is most likely related to the earlier breakdown of the tip vortices as we will be discussing in more detail in section \ref{energy}. The results may indicate that for a `realistic' broadband turbulent inflow, the amplitude of wake meandering do not significantly vary (at least for the FST parameters considered here). }\\

 \begin{figure}
  \centerline{
  \includegraphics[clip = true, trim = 0 0 0 0 ,width= 0.85\textwidth]{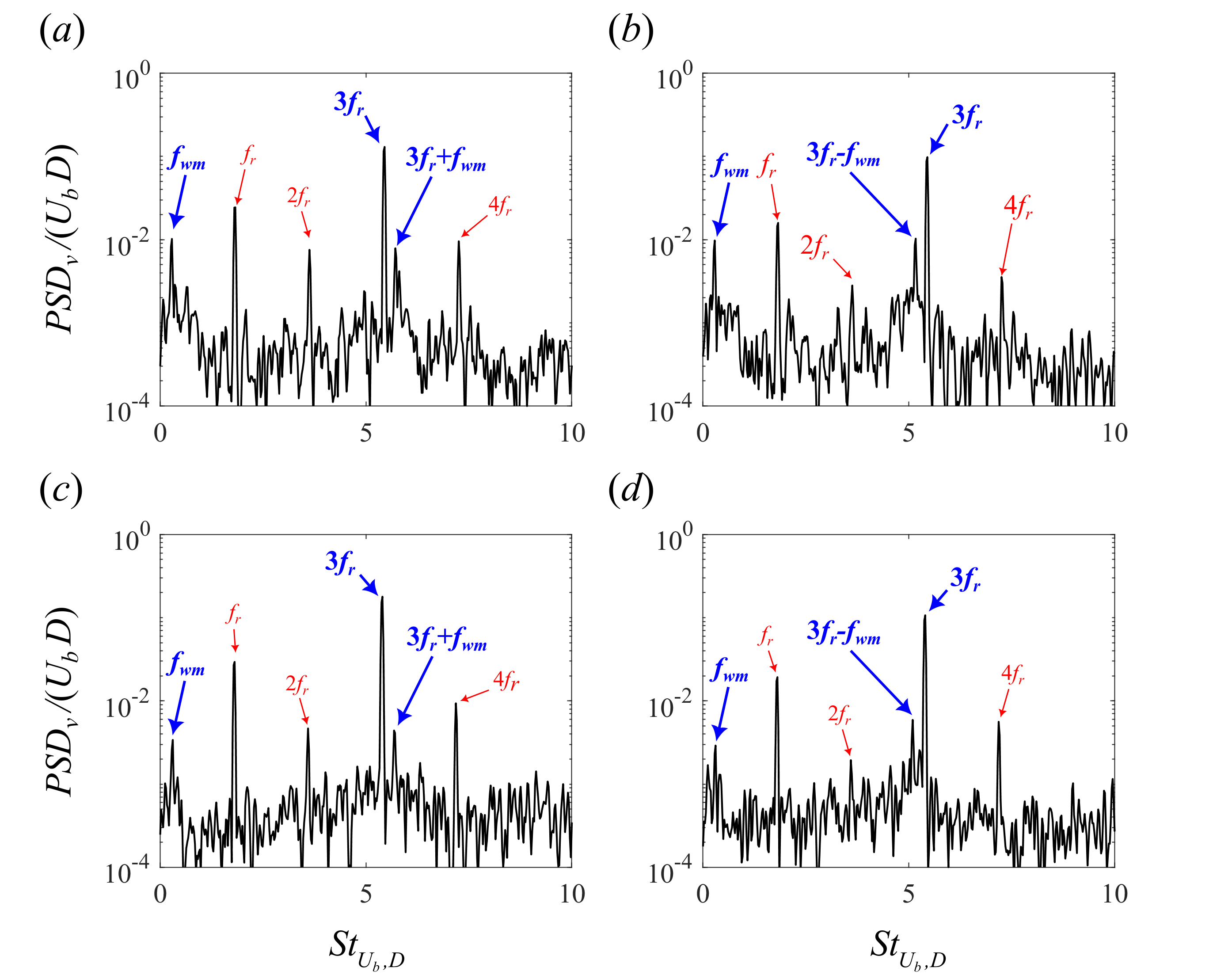}  }
 \caption{Spectra for case 2a at $x/D = 0.5$ and (a) $y/D = 0.55$ and (b) $y/D = 0.52$. Subfigures (c) and (d) show the same for the case 2b. }
\label{fig:43b}
\end{figure}

A closer look at figure \ref{fig:43a}(b), \textit{i.e} that for case 2a at $x/D=0.5$ and $y/D = 0.55$, reveals a new peak at a frequency slightly higher than $3f_r$. To understand the origin of this peak, the spectra are reproduced on a linear scale in figure \ref{fig:43b}(a). This new frequency was found to be numerically equal to $3f_r + f_{wm}$, where $f_{wm} = 0.29$ is the {wake meandering frequency observed towards the outer edge of the wake}. Similarly at $x/D = 0.5$ and $y/D = 0.52$, another new frequency was found which was numerically equal to $3f_r - f_{wm}$. A similar observation, although weaker, was recorded for the case 2b as well as shown in figures \ref{fig:43b}(c-d). {This is an excellent example of interaction between the tip vortices and wake meandering.} More interestingly, the new frequencies are similar to the secondary frequencies observed in previous studies on multiscale flows involving cylinders/prisms of unequal diameters \citep{baj2017interscale, cicolin2021role, biswas2022energy}. The secondary frequencies were numerically equal to the sum and difference of the primary shedding frequencies of the two cylinders and were shown to be produced due to the non-linear triadic interaction between the primary frequencies. Whether the same is true for the new frequencies observed here will be explored in section \ref{energy}.  \\

\begin{figure}
  \centerline{
  \includegraphics[clip = true, trim = 0 0 0 0 ,width= \textwidth]{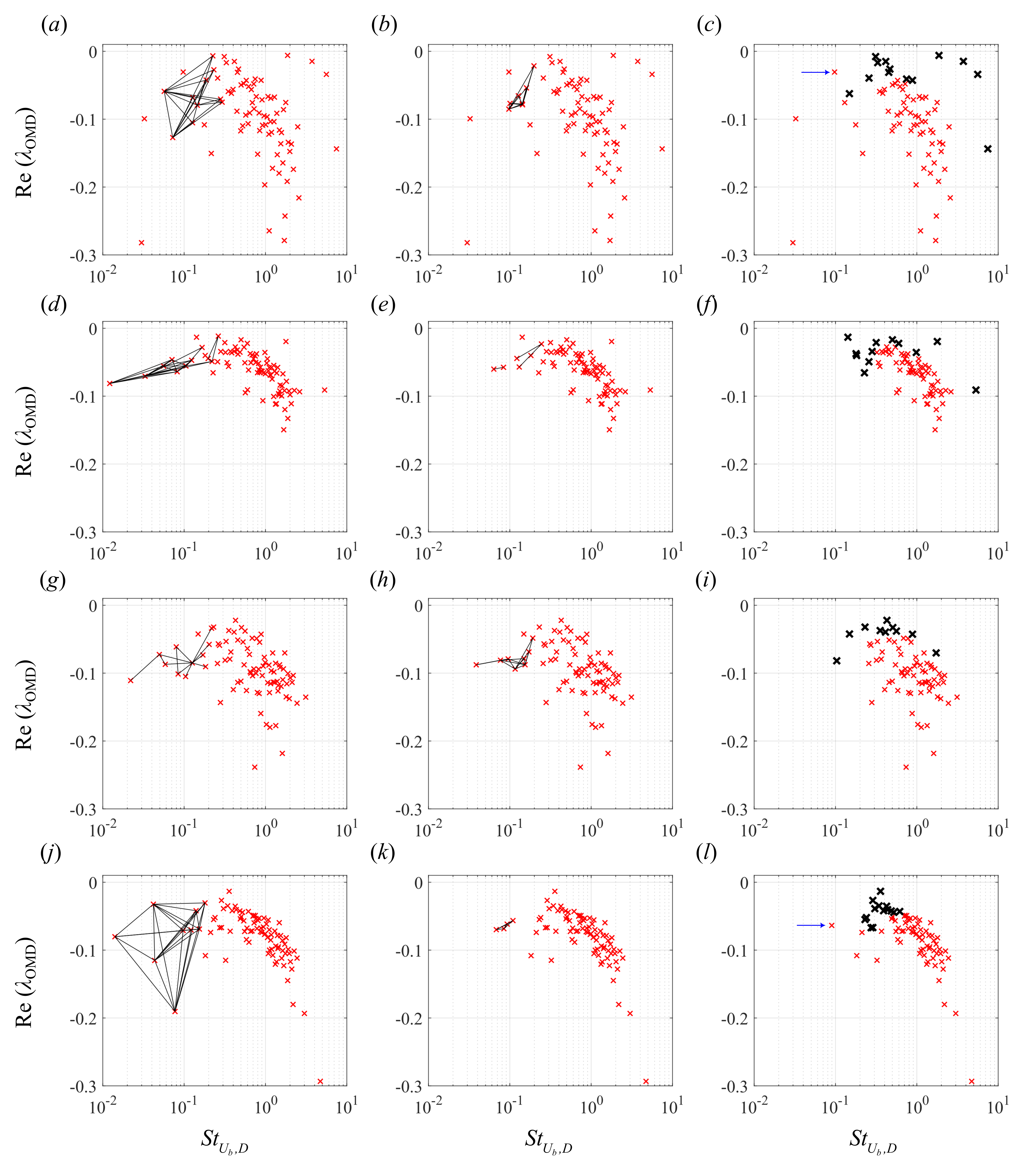}  }
 \caption{ $1^{st}$, $2^{nd}$ and final iterations of the mode clustering algorithm for the no grid case (a-c), case 2c (d-f), case 3b (g-i), and case 4b (j-l). \rfC{The modes indicated by the blue arrows were not selected despite being less damped since the associated mode shapes could not be interpreted as physically meaningful.} }
\label{fig:iter}
\end{figure}

\section{Modal decomposition}

{The coherent motions present in a wind turbine wake have been studied using different modal decomposition techniques such as Proper Orthogonal Decomposition (POD), Dynamic Mode Decomposition (DMD) or some variant of the two \citep{sarmast2014mutual, debnath2017towards, de2021pod, biswas2024energy}. These methods successfully captured the modes related to tip vortices and vortex sheddings from the tower or nacelle and wake meandering \citep{debnath2017towards, de2021pod, kinjangi2023characterization, biswas2024energy}. For the current dataset, we use Optimal Mode Decompoisition (OMD) which is a more generalised version of the originally proposed DMD algorithm \citep{schmid2010dynamic, wynn2013optimal}. \rfB{DMD assumes that the time evolution of the flow can be governed by a time invariant, best-fit linear operator $\boldsymbol{A}$ and the eigen-decomposition of the operator gives the so-called DMD modes. Here the time dynamics are obtained by projecting the flow data onto a POD subspace. Contrastingly, in OMD, a more generalized approach is followed by solving a two-way optimization problem for the a low order subspace of the flow and it's dynamics. \citet{wynn2013optimal} showed that the restrictive nature of DMD can result into significant under performance in some cases compared to OMD \citep{wynn2013optimal}. In our previous work, we compared the performance of OMD with POD in a different flow considering the wake of a multiscale array of bars \citep{baj2015triple}. POD was found to be incapable of correctly identifying the mode shapes further downstream from the wake generating body. Therefore, OMD was the preferred choice. A brief overview of OMD is presented in appendix 2. }

 \citet{biswas2024energy} recently applied OMD to the wake produced by the same wind turbine model as in the present study. Various coherent modes such as the ones associated with the tip vortices, the vortex sheddings from the nacelle and the tower and wake meandering were identified and the effect of the tip speed ratio was discussed. However, the analysis was restricted to the case with negligible freestream turbulence. In this section, we build on the results of \citet{biswas2024energy} by including the effect of freestream turbulence.}

\subsection{Mode clustering}

OMD was performed with a rank $r=175$ to be consistent with our previous work \citep{biswas2024energy}. For brevity, the typical OMD spectra are only shown for four selected cases: 1 (no grid case), 2c, 3b and 4b in figures \ref{fig:iter}(a, d, g, j). A mode with a higher growth rate ($\Re(\lambda_{OMD})$) generally represents a physically meaningful mode. Due to the discrete nature of the tip vortex related frequencies, their identification from the OMD spectra was straightforward. However, it was not so for the low frequency wake meandering related modes. In the vicinity of the wake meandering frequency band (WMFB) \textit{i.e.} in $0.15\leq St_D \leq 0.4$, there were multiple modes with comparable frequencies and growth rates. The associated mode shapes also looked similar. This was more of a problem for the cases with medium to high FST levels. Therefore, in order to simplify the OMD spectra (especially in the vicinity of the WMFB), the spatially (and spectrally) similar modes were identified and grouped together into single representative mode. \rfE{ \citet{beit2021data} proposed a clustering algorithm that grouped/clustered similar OMD modes together to form a smaller set of dissimilar modes. The authors used graph theory to cluster the modes in an iterative process. To assess the similarity of two modes, the authors looked at the similarity between the subspaces spanned by the real and imaginary parts of the complex OMD modes. A constraint based on spectral similarity was also imposed. Finally, two modes were considered to be similar if the spatial and spectral similarity was above a threshold value. The algorithm was later successfully applied to construct the shedding modes behind a cylinder exposed to FST \citep{de2025influence}. We apply the same algorithm for this case, albeit with some changes in the way the similar modes are grouped. The method is discussed in detail in appendix 2. }

 The evolution of the OMD spectra at the first, second and the final iterations of clustering is shown in figure \ref{fig:iter} for the four selected cases. For each case, a sensitivity analysis was performed to select a threshold value/cutoff for spatial and spectral similarity (denoted by $\zeta_{spat}$ and $\zeta_{spec}$ respectively). This is discussed in appendix 2 (please see figure \ref{fig:convergence}). After finding similar modes, they are connected with an `edge' (shown by the black lines). Note that there exists a number of similar modes only in the vicinity of the WMFB. Once the similar modes are identified, the problem of clustering becomes equivalent to finding subgroups/clusters where each mode is connected with every other mode with an edge. In addition, a subgroup should not be strictly contained in another group. In order to obtain the mode representing a cluster, the modes in the same cluster are multiplied with a weight (related to the growth rate of the modes) and stored in a matrix. Thereafter, the first singular value of the matrix was simply used to represent the mode associated to the cluster. Please refer to appendix 2 for further details.

Interestingly, a significant number of similar modes are observed for the no grid case as well. Note that, the recent work by \citet{biswas2024energy} applied OMD on the same data set as case 1 here without mode clustering. {The modes obtained with/without mode clustering were similar for the no grid case as will be discussed in the next paragraph. Mode clustering therefore simplified} the spectrum in the present case, making it easier to select the physically meaningful modes which are used to form a reduced order representation of the coherent component of the flow. {Further details about the mode selection process can be found in appendix 3. }

 \begin{figure}
  \centerline{
  \includegraphics[clip = true, trim = 0 0 0 0 ,width= \textwidth]{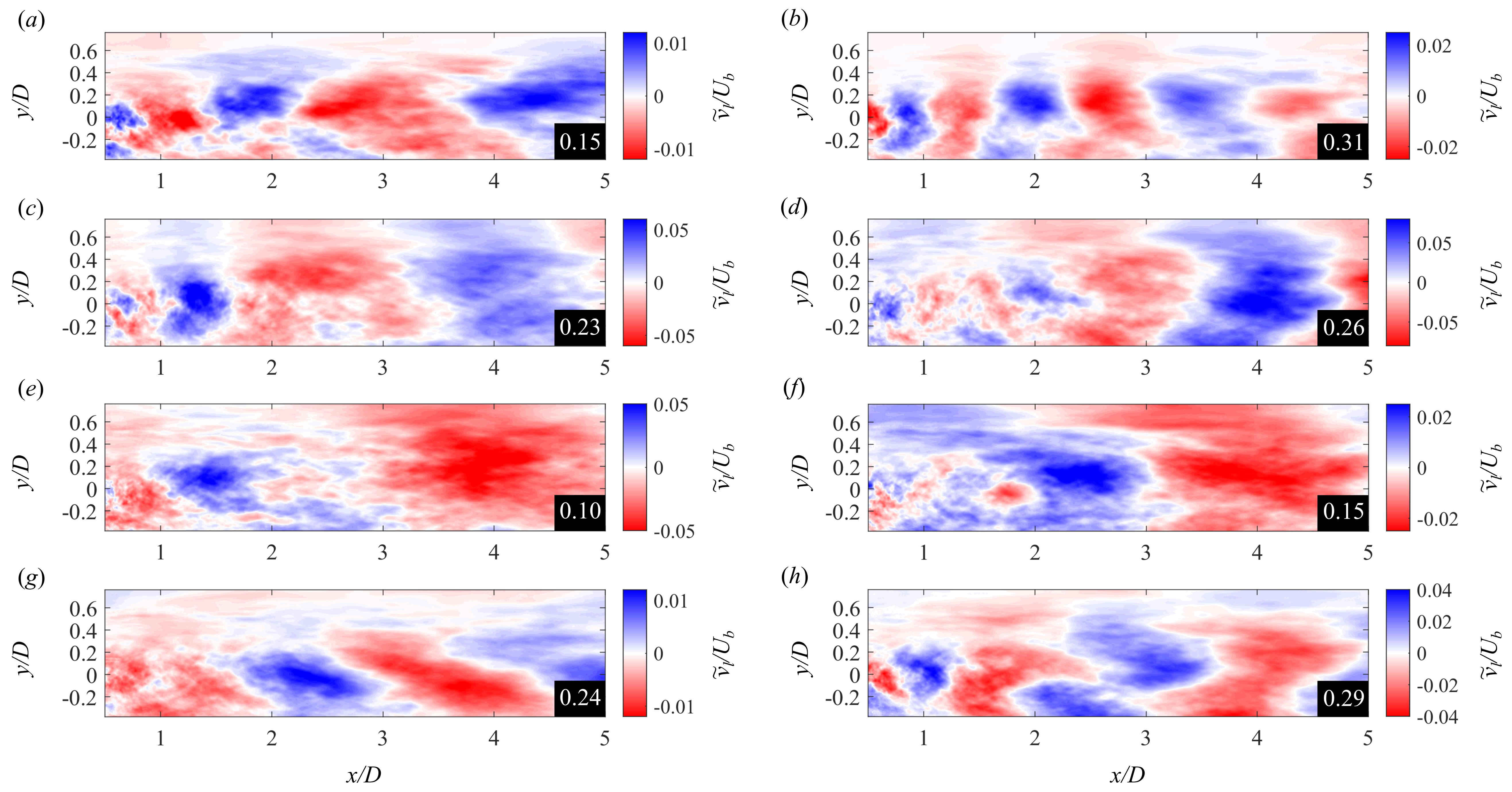}  }
 \caption{ Two dominant wake meandering modes for the cases (a,b) 1, (c,d)  2c, (e,f) 3b, (g,h) 4b. The associated Strouhal numbers ($St_{U_b,D}$) are also highlighted in the figures.  }
\label{fig:WM_FST}
\end{figure}

Upon clustering, the selected modes are highlighted in figures \ref{fig:iter}(c, f, i, l) with the `$\boldsymbol{\times}$' symbols. Among the tip vortex related modes, only the frequencies $f_r - 4f_r$ could be captured in the OMD spectrum (figure \ref{fig:iter}(c)). \rfA{The higher harmonics, $5f_r$ and $6f_r$ are weaker in nature and are concentrated close to the rotor. \citet{biswas2024energy} could not detect these frequencies in the OMD spectra even with a high rank ($r=175$) of the OMD matrices. The modes corresponding to these frequencies were therefore obtained using phase averaging. The snapshots were divided into 48 phase bins (with phases ranging from $-\pi$ to $\pi$) at the reference frequency. Subsequently, the second Fourier mode of the phase-averaged flow field was taken in order to produce a complex spatial mode that mimics the OMD modes \citep{baj2017interscale, biswas2022energy}.} For all cases, the final modes (marked by the `$\boldsymbol{\times}$' symbols) were carefully selected by looking at the spatial mode shape and how the clusters evolved over iterations. The time varying coefficients of the modes were obtained by projecting the snapshots onto the selected modes.

\subsection{The nature of the wake meandering mode in the presence of FST}

\rfB{The transverse velocity component ($\tilde{v}_l$, where $l$ runs over the coherent modes)} of the two dominant wake meandering modes (having a frequency in the WMFB) for the four cases of increasing $T_i$ are shown in figure \ref{fig:WM_FST}. For the no grid case (figure \ref{fig:WM_FST}(a,b)), regular patterned structures of wavelength $1.5-2D$ are observed. \rfC{The modes start from close to the nacelle in the inner part of the wake and expand further downstream.} Note that, these two modes are almost exactly the same as the modes reported by \citet{biswas2024energy} without mode clustering (see figures 4(a,b)), apart from a slight phase shift and a slight difference in the mode amplitude. This offers reassurance that the mode clustering algorithm adopted here does not alter the physical information embedded in the data set, rather it simplifies the analysis. {For case 2c (low $T_i$), the dominant wake meandering modes in figures \ref{fig:WM_FST}(c,d) still show similar regular structures, but the modes exhibit a larger transverse (along $y$) width compared to that for the no grid case.} Note from figure \ref{fig:WM_FST}(a,b) that the modes {for the no grid case} are not particularly energetic for $y\geq 0.5D$. The modes initiate from the vicinity of the nacelle highlighting the importance of the nacelle to {providing} the perturbation necessary to kick start wake meandering \citep{howard2015statistics, biswas2024effect}. For higher $T_i$, the modes show a significant energy content for $y \geq 0.5D$ {as was discussed in section \ref{near}}. They also become increasingly irregular in nature with increasing FST levels. A higher transverse width of the modes can however still be observed, at least for the medium $T_i$ case (case 3b).}

The energy budget analysis in \citet{biswas2024energy} revealed that in the absence of FST, the wake meandering modes were primarily energised by the mean flow. In the presence of FST, whether the dominant energy source for the mode stays the same or not is an interesting question and it will be discussed further in the following sections.

 \begin{figure}
  \centerline{
  \includegraphics[clip = true, trim = 0 0 0 0 ,width= \textwidth]{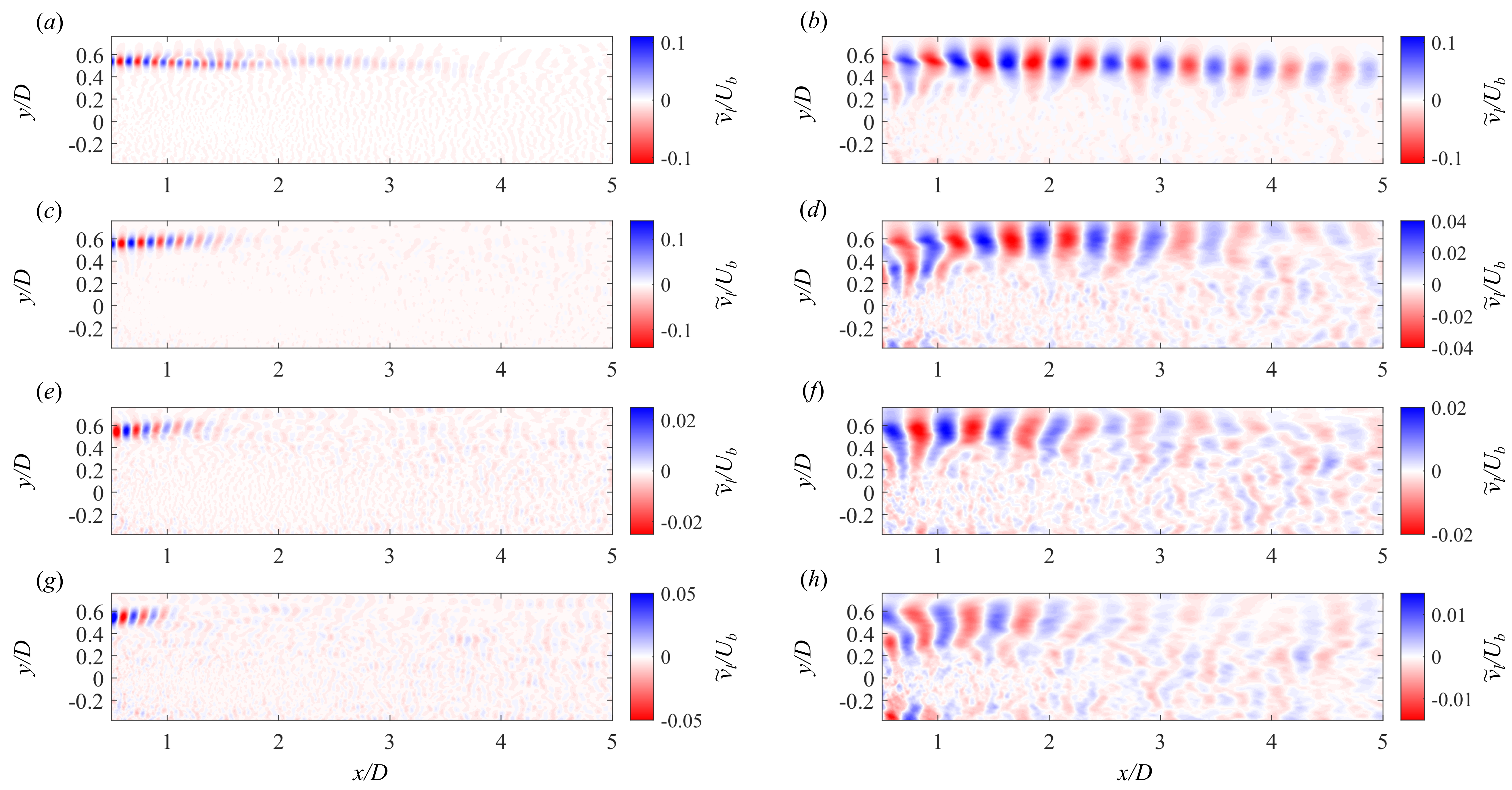}  }
 \caption{ Modes associated with $3f_r$ (a,c,e,g) and $f_r$ (b,d,f,h) for the no grid case (a,b), case 2c (c,d), case 3b (e,f) and case 4b (g,h).  }
\label{fig:tip_FST}
\end{figure}

 \begin{figure}
  \centerline{
  \includegraphics[clip = true, trim = 0 0 0 0 ,width= 0.8\textwidth]{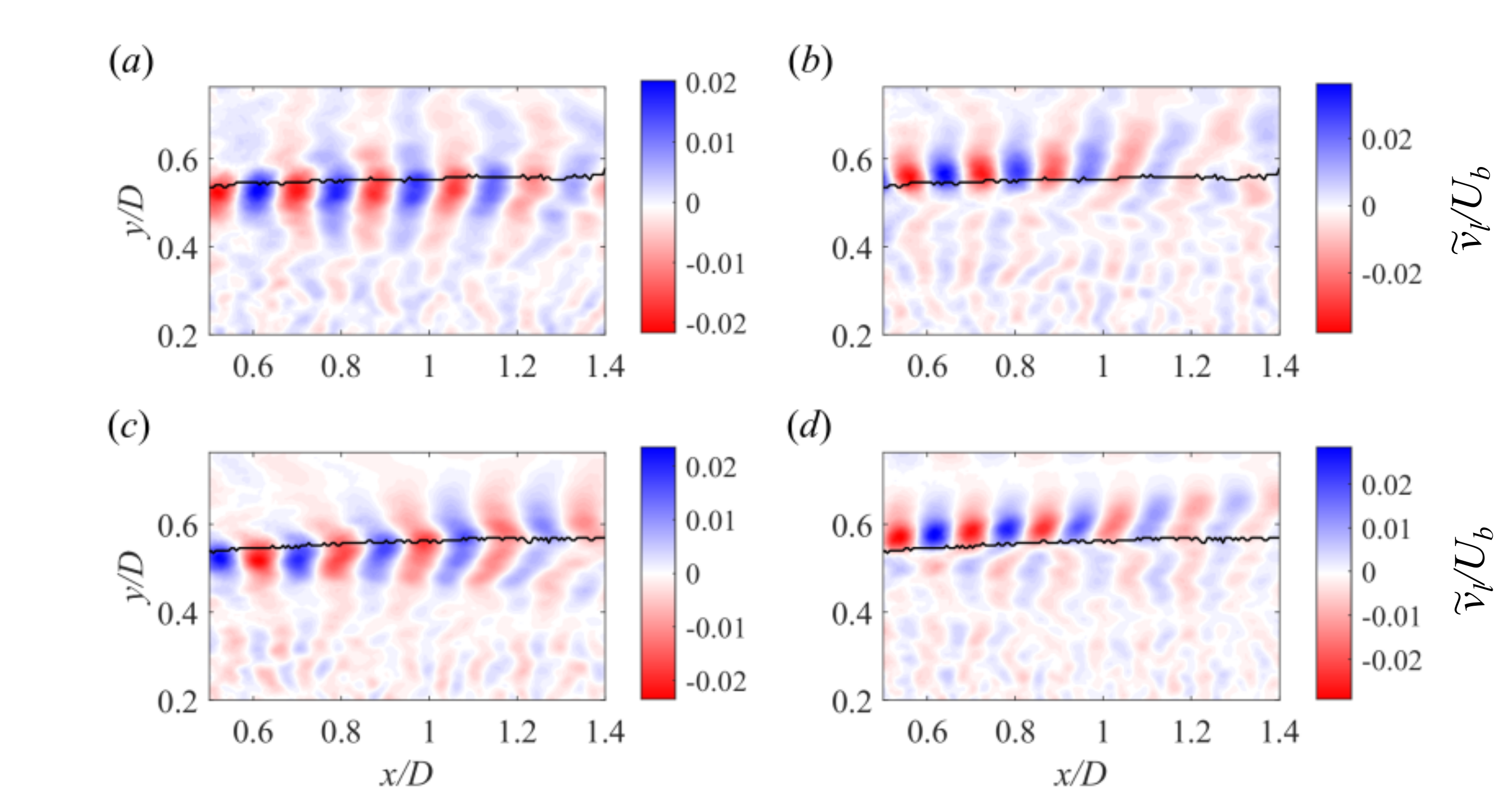}  }
 \caption{   Modes associated with $3f_r-f_{wm}$ (a,c) and $3f_r+f_{wm}$ (b,d) case 2a (a,b) and case 2b (c,d).  }
\label{fig:secondary_modes_fst}
\end{figure}

\subsection{The modes related to the tip vortices}

\rfD{Among the tip vortex related frequencies, only the modes associated with $3f_r$ and $f_r$ are shown in figure \ref{fig:tip_FST} for the same four cases (1, 2c, 3b, 4b) since they were the most important and were observed for all the cases (see figure \ref{fig:43a})}. The modes associated with $3f_r$ clearly show how the tip vortices break down earlier as freestream turbulence intensity is increased. A similar pattern is observed for the modes associated with $f_r$ as well as the harmonics of $f_r$ and $3f_r$ (not shown for brevity). Furthermore, the modes with FST appear to have a larger transverse spread. This is likely due to the vortex wandering or vortex jittering reported in previous studies in the presence of FST \citep{gambuzza2023influence}.

{Note that for the cases 2a and 2b, new frequencies were observed that corresponded to the sum and difference of $3f_r$ and a wake meandering frequency that was dominant near the edge of the wake}. These frequencies were however not present in the OMD spectrum for either of the cases. {A similar observation was reported in our previous studies where the new frequencies corresponding to the sum and difference of the dominant coherent modes were not captured by OMD \citep{baj2017interscale, biswas2022energy}}. These modes therefore had to be obtained using phase averaging. These are shown in figure \ref{fig:secondary_modes_fst} for cases 2a and 2b. The spatial orientation of the modes reveals an interesting pattern. Note from figure \ref{fig:freestream_wake_fft}(b) that the $f_{wm}$ mode interacting with $3f_r$ was energetic in the freestream side ($y>0.5D$). In order to indicate the difference in the spatial location of the modes, a black line was added to figure \ref{fig:secondary_modes_fst} showing the loci of the points where the kinetic energy of the mode $3f_r$ was maximum. It can be seen that for both the cases, the high frequency mode $3f_r + f_{wm}$ is situated more towards the freestream side, while $3f_r - f_{wm}$ resided more towards the wake side. \rfE{The exact same spatial orientation was observed for secondary frequencies in fundamentally different multiscale flows \citep{biswas2022energy, baj2017interscale}. \citet{biswas2022energy} considered the wake of a cylinder (main cylinder) and control rod (another cylinder of diameter $1/10{th}$ of the main cylinder). Apart from the shedding modes associated with the main cylinder ($f_m$) and control rod ($f_c$), secondary frequencies $f_c\pm f_m$ were observed. The high frequency mode resided more towards the shear layer between the two wake generating bodies, while the low frequency mode resided in the other outer side of the wake of the smaller wake generating body (having a higher shedding frequency). A similar observation was reported by \citet{baj2017interscale} in the wake of a multiscale array of prisms. These secondary frequencies were shown to be a result of non-linear traidic interaction between the two primary shedding frequencies}. The fact that we observe such similarity between simple two dimensional multiscale flows and a complex three dimensional multiscale flow involving rotation is truly remarkable and adds credance to the notion {that these multiscale interactions are indeed universal}.

 \begin{figure}
  \centerline{
  \includegraphics[clip = true, trim = 0 0 0 0 ,width= 1\textwidth]{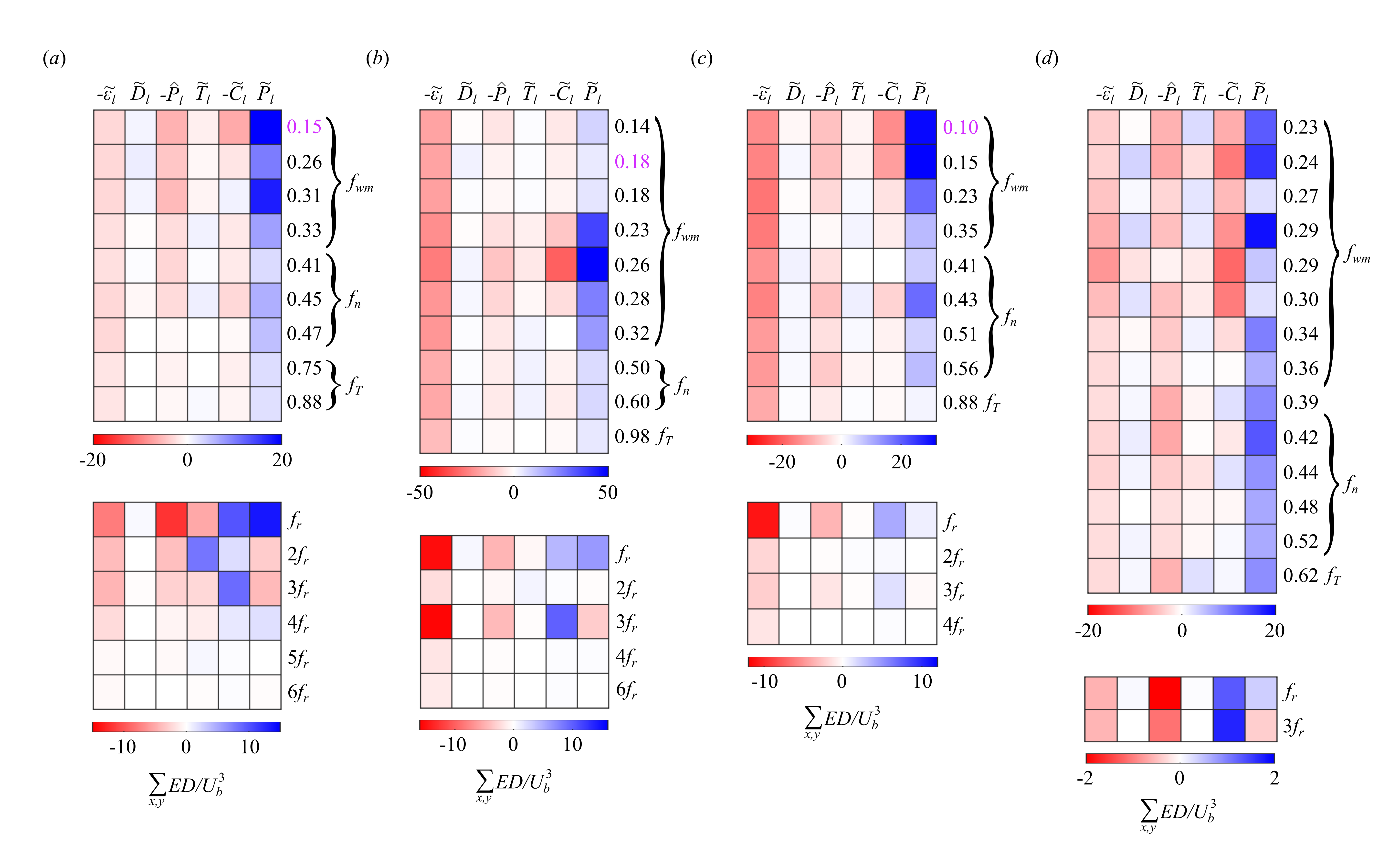}  }
 \caption{ Summed budget terms for (a) no grid case, (b) case 2c, (c) case 3b and (d) case 4b. \rfD{The numbers on the right show the associated Strouhal numbers ($St_{U_b,D}$) of the modes. For the clustered modes, the $St_{U_b,D}$ values are highlighted in purple.} $E$ is a generic variable representing the budget terms. }
\label{fig:budget_fst}
\end{figure}

\section{Energy exchanges}
\label{energy}

In this section, we use the {multi-scale triple-decomposed} coherent kinetic energy budget equation derived by \citet{baj2017interscale} to assess the energy budgets of the individual coherent modes. \rfB{\citet{baj2017interscale} derived separate equations for the kinetic energy associated with the mean flow ($\bar{k}$), each of the coherent modes ($\tilde{k}_l$ for the $l_{th}$ coherent mode) and the stochastic component of the flow ($k^{''}$). For the present work, we only use the equation for the coherent modes which can be symbolically represented as the following:}

\begin{equation}
    \textcolor{black}{\frac{\partial \tilde{k}_l}{\partial t}   = -\textcolor{black}{\tilde{C}_l} + \tilde{P}_l - \textcolor{black}{\hat{P_l}} +\textcolor{black}{\Big( \tilde{T}^+_l - \tilde{T}^-_l\Big)} -  \textcolor{black}{\tilde{\epsilon_l}} +\textcolor{black}{\tilde{D_l} }}
    \label{eqn:coherent_TKE}
\end{equation}

Here, the source terms on the right hand side consist of convection {($\tilde{C}_l$)}, mean flow production ($\tilde{P}_l$) or energy transfer from the mean flow to the $l_{th}$ coherent mode, production of stochastic turbulent kinetic energy directly from the coherent mode $l$ ($\hat{P}_l$), triadic energy production ($\tilde{T}^+_l - \tilde{T}^-_l$), direct dissipation from coherent mode $l$ (${\tilde{\epsilon}}_l$) and diffusion ($\tilde{D}_l$). \rfB{Note that all the terms in equation \ref{eqn:coherent_TKE} are averaged over time.} \citet{baj2017interscale} showed that the triadic energy production term can become significant only when there exist three frequencies that linearly combine to zero or, in other words, there is a triad in the form \rfE{$f_l \pm f_s \pm f_t = 0$}. The constituents of this term were defined as follows:

\begin{equation}
    \tilde{T}^+_{l} = -\frac{1}{2}\sum_{f_s,f_t}\overline{\tilde{u}^{f_l}_i \tilde{u}^{f_t}_j \frac{\partial \tilde{u}^{f_s}_i}{\partial x_j}}, \hspace{1cm} \tilde{T}^-_{l} = -\frac{1}{2}\sum_{f_s,f_t}\overline{\tilde{u}^{f_s}_i \tilde{u}^{f_t}_j \frac{\partial \tilde{u}^{f_l}_i}{\partial x_j}}.
    \label{eqn:triad}
\end{equation}

Here, $\tilde{T}^+_{l}$ denotes the nonlinear energy transfer from $f_s$ to $f_l$, while $\tilde{T}^-_{l}$ denotes the energy lost by $f_l$ to $f_s$. In both these terms, $f_t$ acts as a mediator or a catalyst \citep{baj2017interscale, yeung2024revealing}. \rfB{$\tilde{u}^{f_l}_i$ denotes the $i_{th}$ velocity component of the coherent mode $l$ (having frequency $f_l$).} The total triadic energy gain/loss for $f_l$ is obtained by summing over all the frequencies ($f_s$, $f_t$) that form a triad with $f_l$. The other term that is of particular interest is the $\tilde{P}_l$ term since it is directly linked to wake recovery. $\tilde{P}_l$ is defined as $\tilde{P}_l = -\sum_m \overline{\tilde{u}^m_i \tilde{u}^l_j} \frac{\partial \overline{u}_i}{\partial x_j}$, where $m$ can be any coherent mode selected in the reduced order representation of the flow including the $l_{th}$ mode. The complete composition of {each of the terms of equation \ref{eqn:coherent_TKE}} can be found in \citet{baj2017interscale}.

{\citet{biswas2024energy} recently used this equation to obtain the energy budget terms of each of the individual coherent modes in a wind turbine wake with negligible freestream turbulence.} It was reported that the modes associated with the nacelle/tower's vortex shedding and wake meandering were energised by the mean flow. Contrastingly, the tip vortex related modes could {be energised through different pathways} such as energy transfer from the mean flow or non-linear triadic interactions {between various coherent modes co-existing in the wake}. The role of the triadic interactions in the tip vortex stability and merging process was quantified for the first time. \citet{yeung2024revealing} recently proposed a Triadic Orthogonal Decomposition (TOD) which identified coherent structures that optimally account for triadic interactions and quantified the triadic energy transfers. The authors used the same PIV dataset obtained by \citet{biswas2024energy} and reported that the energy transfers obtained using the new method were in good agreement with the previous work. Although these studies have significantly improved our understanding of the origin and evolution of the various coherent structures in a wind turbine wake, the effect of freestream turbulence was not considered in these studies. {This section attempts to fill that gap}.   \\


 \begin{figure}
  \centerline{
  \includegraphics[clip = true, trim = 0 0 0 0 ,width= \textwidth]{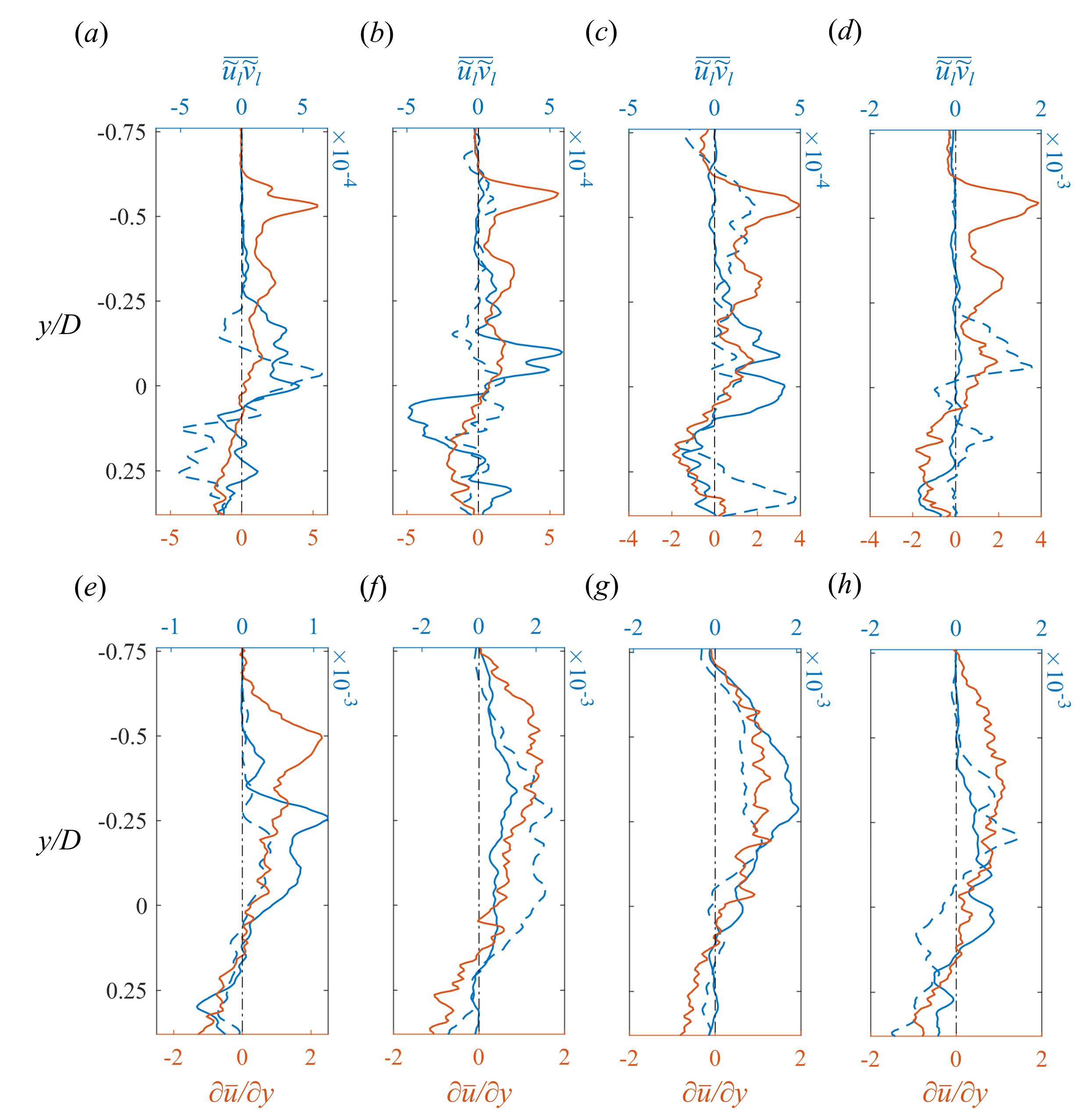}  }
 \caption{ {Profiles of normalised coherent velocity  ($\overline{\tilde{u}^l \tilde{v}^l}$) and transverse gradient of mean streamwise velocity ($\frac{\partial \overline{u}}{\partial y}$) for the wake meandering modes extracting the highest amount of energy from the mean flow for the no grid case (a, e), case 2c (b, f), case 3b (c, g) and case 4b (d,h). The profiles are shown at $x/D=1$ (a-d) and $x/D=4$ (e-h). } }
\label{fig:corr_grad}
\end{figure}

The kinetic energy budget terms are first discussed for the four cases {analysed} previously. The energy budget terms are summed over the entire domain and shown in figure \ref{fig:budget_fst}. Since the modes related to the tip vortices show weaker energy exchanges compared to the low frequency modes, a separate colour bar is used for them. On the right, the numbers show the Strouhal numbers ($St_{U_b,D}$) associated with the modes. The modes are also classified as either related to wake meandering ($0.15\leq f_{wm} \leq 0.4$), nacelle's shedding ($0.4\leq f_n\leq 0.6$) or the tower's shedding ($0.6\leq f_T \leq 1$) as discussed in section \ref{wm}. On some occasion this classification could not be made strictly based on the frequency, particularly for the clustered modes. As already discussed, for these modes, the frequency associated with the cluster was obtained as a weighted average of the frequencies of the modes forming the cluster (which may differ from the actual frequency of the mode). Therefore, the clustered modes were carefully classified by looking at their mode shape and the frequency content in the associated time varying mode coefficient. An example can be seen in figure \ref{fig:budget_fst}(c), where $St_{U_b,D}=0.1$ was classified as related to wake meandering.   


\citet{biswas2024energy} showed that in the presence of negligible freestream turbulence, the wake meandering mode acted like a `primary mode', extracting energy from the mean flow. In the presence of FST, the mean flow production ($\tilde{P}_l$) term is still found to be the dominant energy source for wake meandering. {The $\tilde{P}_l$ terms for the dominant wake meandering modes in the presence of FST were always found to be stronger than that for the no grid case which is in line with the discussion in section \ref{near}. Furthermore, the dominant wake meandering modes for case 2c drew the highest amount of energy from the mean flow. This is in line with the highest far wake meandering amplitude seen for this FST case in figure \ref{fig:a_wm_avg}. The $\tilde{P}_l$ term then became relatively weaker as $T_i$ was increased further, likely due to the wake meandering modes becoming less coherent (as observed in figure \ref{fig:WM_FST}) and less efficient at extracting energy from the mean flow. This was accompanied by an enhancement in the relative contributions from the triadic interaction and diffusion term for the highest $T_i$ case.}


 The tip vortex related modes in the presence of FST showed energy exchanges similar to the no grid case, however, the magnitudes of the terms decreased as $T_i$ was increased. $f_r$ for the no grid case was primarily energized by the mean flow production ($\tilde{P}_l$) term. For the FST cases, the relative contribution of the convection term for $f_r$ was higher than the $\tilde{P}_l$ term. As $T_i$ is increased, the location where the kinetic energy of the $f_r$ mode is maximum moves further upstream. Therefore a larger proportion of the $\tilde{P}_l$ is not captured within the {PIV field of view and eventually this missing energy is `picked up'} by the convection term. A field of view large enough to capture the full spatial growth and decay of the mode would yield zero total convection. Note that the tip vortices ($3f_r$) are formed due to the difference in pressure between the pressure and suction surface of the blades \citep{andersen2013simulation}. {$3f_r$ is therefore energised upstream of the FOV and simply decays (due to different energy transfers) within the FOV under consideration. }Interestingly, the $2f_r$ mode was found to be energized by the triadic interaction term for all the FST cases in which it could be captured. A similar observation was reported by \citet{biswas2024energy} when FST was negligible. $2f_r$ was therefore identified as a secondary mode that is produced purely due to non-linear interaction at $\lambda=6$. The magnitude of the triadic interaction term diminished drastically with $T_i$ indicating weaker non-linear interaction. For the highest $T_i$ cases, the harmonics of $f_r$ and $3f_r$ were not observed within the field of view and hence no triadic interaction could be captured.

{The intriguing question is why is the $\tilde{P}_l$ term at all stronger for the wake meandering modes in the presence of FST? To answer this we need to look at the composition of the $\tilde{P}_l$ term. As discussed in \citet{biswas2024energy}, utilizing the fact that the transverse gradient of streamwise mean velocity ($\frac{\partial \bar{u}}{\partial y}$) is much stronger than the other mean velocity gradients and assuming that the velocity components of the different modes are uncorrelated ($\overline{\tilde u^l \tilde v^n} \approx 0$), we can approximate the mean flow production term as $\tilde{P}_l \approx -\overline{\tilde{u}^l \tilde{v}^l} \frac{\partial \overline{u}}{\partial y}$. This was verified to be true for all the modes and is not shown here for brevity, a more detailed discussion can be found in \citet{biswas2024energy}. \rfE{In figure \ref{fig:corr_grad}, the profiles of $\overline{\tilde{u}^l \tilde{v}^l}$ and $\frac{\partial \overline{u}}{\partial y}$ are shown for the two important wake meandering modes (as shown in figure \ref{fig:WM_FST}) for the four cases (1, 2c, 3b and 4b) at two streamwise locations, $x/D=1$ (representative of near wake) and $x/D=4$ (representative of far wake). Note that, the no grid case shows the strongest gradients in the tip shear layer. However the wake meandering modes remain sheltered well within the shear layer (see figure \ref{fig:WM_FST}(a,b)) in the near wake as well as the far wake, inhibiting it from utilizing the stronger gradients available in the shear layer. For the other cases with FST, in the near wake, the wake meandering modes still remain sheltered, explaining the similar wake meandering amplitudes observed for all the cases in the near wake (figure \ref{fig:a_wm_avg}). Further downstream, due to the earlier breakdown of the tip vortices, the modes extend beyond the shear layer (see figure \ref{fig:WM_FST}). As a result, although the mean velocity gradients are slightly weaker (due to enhanced turbulent mixing), the modes do a better job of exploiting this shear thereby yielding a higher rate of energy transfer from the mean flow. } }

 \begin{figure}
  \centerline{
  \includegraphics[clip = true, trim = 0 0 0 0 ,width= \textwidth]{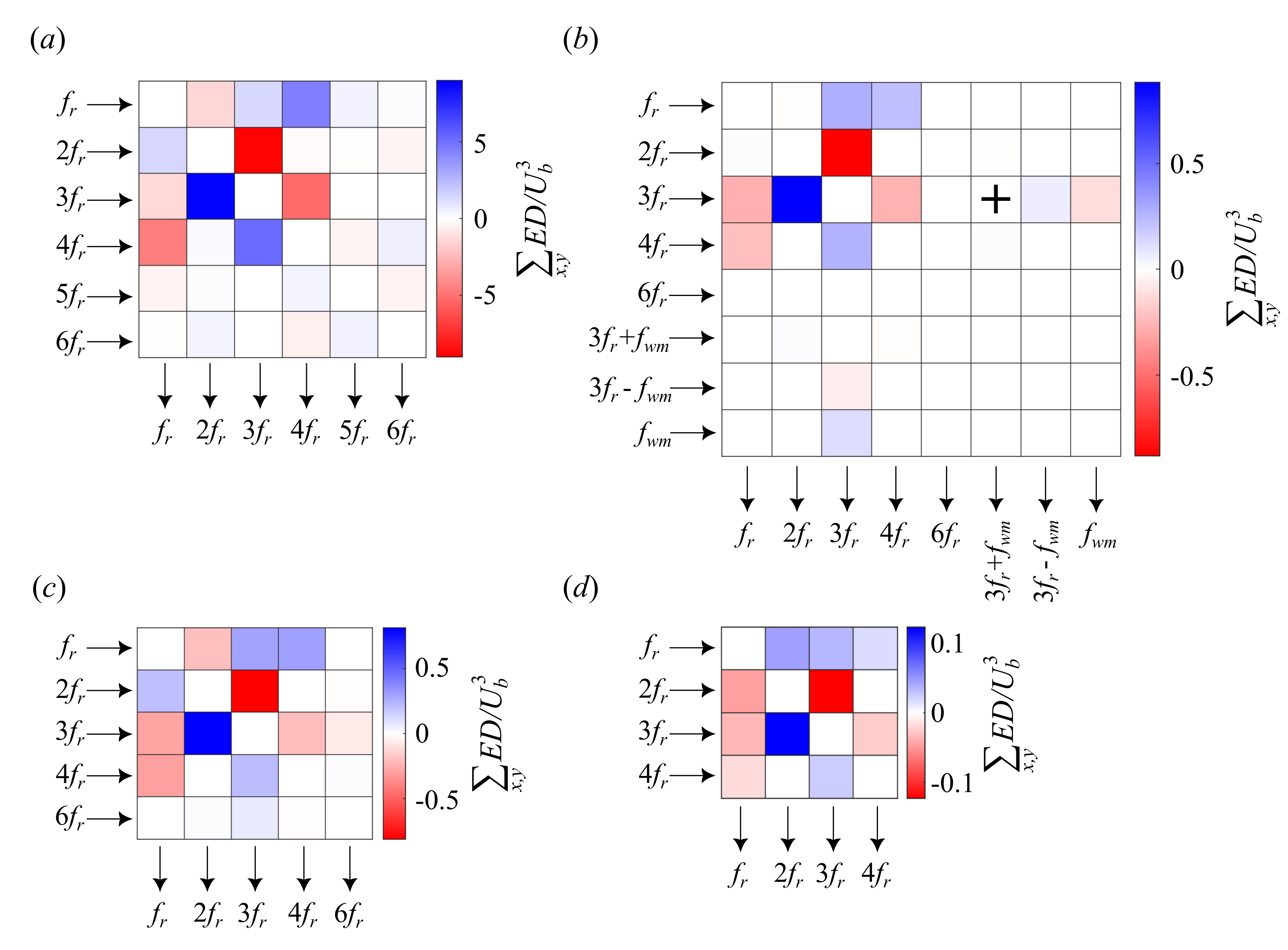}  }
 \caption{ Inter frequency triadic energy transfers for (a) no grid case, (b) case 2a, (c) case 2c and (d) case 3b.  }
\label{fig:triad_fst}
\end{figure}

 \begin{figure}
  \centerline{
  \includegraphics[clip = true, trim = 0 0 0 0 ,width= \textwidth]{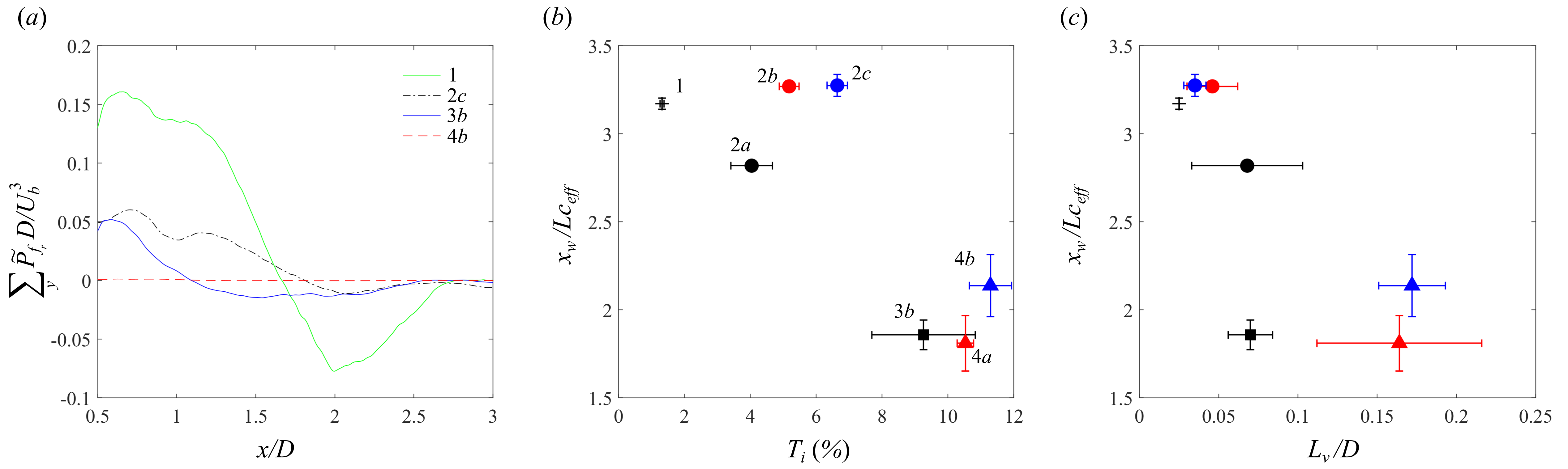}  }
 \caption{ (a) Streamwise variation of spanwise averaged mean flow production term for $f_r$ for the different FST cases. \rfD{The variation of the wake recovery onset location ($x_w$) is shown against (b) $T_i$ and (c) $L_v$. $x_w$ is scaled by the convective length scale ($Lc_{eff} = \pi D/\lambda_{eff}$). } }
\label{fig:recovery_FST}
\end{figure}

\subsection{Triadic interactions}

In equation \ref{eqn:triad}, if we only sum over $f_t$, we can obtain the triadic energy transfer between $f_l$ and $f_s$. \citet{biswas2024energy} reported these inter-frequency triadic energy transfers in the tip vortex system (formed by the modes $f_r - 6f_r$) for negligible $T_i$ and found the same pattern of triadic energy transfers for a range of $\lambda$s ($\lambda \geq 5.5$). Furthermore, the triadic energy transfers were mainly concentrated within the first four frequencies $f_r-4f_r$. We perform a similar analysis here for different FST levels. The inter-frequency triadic energy transfers (summed over the entire domain) are obtained for select cases (1, 2a, 2c, 3b) and shown in figure \ref{fig:triad_fst}. Note that no triads were formed for case 4 as $f_r$ and $3f_r$ were the only two frequencies detected in the FOV. For the no grid case, the energy exchange pattern is the same as that observed for $\lambda=6$ in \citep{biswas2024energy}. \rfC{\citet{biswas2024energy} identified two important energy transfer pathways: one from $3f_r$ to $f_r$ via $2f_r$, other from $f_r$ to $3f_r$ via $4f_r$. The former is clearly associated with the tip vortex merging process. The second pathway was considered as a feedback that slowed down the energy drainage of $3f_r$.} \rfB{Interestingly, the same pattern is observed for the low turbulence intensity cases (2a, 2c). However, the transfers were an order of magnitude weaker.} It is worthwhile mentioning that for case 2a, the modal decomposition and the energy budget analysis were performed on a truncated domain $x\leq 3D$ in order to avoid boundary layer effects as discussed previously. Nevertheless, it did not impact the values of the total triadic energy fluxes for this case, since the fluxes decayed much before $x=3D$. From figure \ref{fig:triad_fst}(b) it can be seen that the new frequencies $3f_r \pm f_{wm}$ are energised by $3f_r$, \textit{i.e.} the mode having the higher frequency among the two main interacting modes. This was not visually clear for $3f_r \pm f_{wm}$, so a $\boldsymbol{+}$ sign is added in figure \ref{fig:triad_fst}(b). This {energy pathway} is similar to that observed for the secondary modes in the simple multiscale flow discussed in \citep{biswas2022energy}. A similar energy exchange pattern is observed for case 3 (figure \ref{fig:triad_fst}(d)), but the transfer magnitudes were much weaker {in comparison} to the no grid case. \rfE{These results quantitatively indicate that freestream turbulence weakens the tip vortices by weakening the non-linear triadic energy exchanges, although the fundamental nature of the triadic interactions remain the same.}

\subsection{Wake recovery}
The tip vortices in the near wake act as a shield, restricting the exchange of momentum with the freestream \citep{medici2005experimental, lignarolo2015tip} {therefore}, it is essential to break them down to initiate the wake recovery process. To understand the link between the tip vortex breakdown process and wake recovery, \citet{biswas2024energy} studied the mean flow production term ($\tilde{P}_l$ in equation \ref{eqn:coherent_TKE}) of the tip vortex system (consisting of the frequencies $f_r - 6f_r$). It was reported that the tip vortex system exchanges energy with the mean flow primarily through the turbine's rotational frequency $f_r$. Nearer to the turbine, the $f_r$ mode extracted energy from the mean flow ($\tilde{P}_{f_r}>0$) and after some distance downstream, it transferred some energy back to the mean flow ($\tilde{P}_{f_r}<0$) therefore marking the onset location of wake recovery (henceforth denoted by $x_w$). This sign change in $\tilde{P}_{f_r}$ was shown to be due to a change in the inclination of the vortex triplet (having a frequency $f_r$) produced by merging \citep{biswas2024energy}. A similar conclusion was reached by \citet{lignarolo2015tip} for a two bladed turbine.

In figure \ref{fig:recovery_FST}(a), we show the streamwise evolution of $\tilde{P}_{f_r}$, summed along $y$, for four cases of increasing $T_i$. Note that, a sign change in $\tilde{P}_{f_r}$ is observed even with freestream turbulence. The streamwise location where it occurs one an average moves upstream as freestream turbulence intensity is increased. This clearly shows how the wake recovers more quickly in the presence of freestream turbulence. Figures \ref{fig:recovery_FST}(b-c) show how the wake recovery onset location ($x_w$) changed with $T_i$ and $L_v$ respectively for all the cases for which it could be measured. For the cases with the highest $T_i$s, the $\tilde{P}_{f_r}$ term was negligible, at least within the field of view considered, hence, $x_w$ could not be measured. Recently, \citet{biswas2024effect} introduced a new length scale (termed as the convective pitch, $Lc = \pi D/\lambda$) which could be physically interpreted as the distance traveled by a fluid parcel in the freestream in the time taken by the turbine to complete a single revolution. \rfE{Note that $Lc$ is not exactly the same as the pitch of the tip vortices as the tip vortices propagate at a velocity slightly smaller than the freestream velocity.} In figure \ref{fig:recovery_FST}(b-c), $x_w$ is presented in terms of $Lc{_{eff}}$ which is obtained based on $\lambda_{eff}$ for each case. \rfB{Apart from the fact that $x_w$ generally reduces with increasing FST, it appears that for low $T_i$, $x_w$ scales better with $L_v$.  Although figure \ref{fig:recovery_FST} lacks data for higher $L_v$, {it is seen that for the highest $L_v$ for which $x_w$ could be calculated, $x_w$ clearly did not follow the decreasing trend observed for lower $L_v$. This indicates that there is could be a threshold length scale for which the initiation of wake recovery is the closest to the turbine. Furthermore, this threshold length scale was close to the average size of the tip vortices near the turbine. The size of the tip vortices was calculated by fitting a circle to the phase averaged vorticity field of the tip vortices ($3f_r$) for the no grid case (not shown here). The best fit circle had a diameter $\approx 0.07D$, close to where the trend in figure \ref{fig:recovery_FST}(c) changes. This observation is similar to that of \citet{gambuzza2023influence} who noted that for very large integral length scales in a non-Kolmogorov type flow, the near wake length increased back to a value close to the non-turbulent case. Note that the near wake length is loosely connected to the wake recovery location defined here as both depend on the stability of the tip vortices. Another recent study by \citet{bourhis2025impact} looked at the frequency dynamics in the near wake of a model wind turbine exposed to FST. The smallest integral length scales the authors tested were comparable to the largest studied here. Unsurprisingly, they did not find any clear influence of the integral length scales on the strength of the leading tip vortex frequencies. }}

\section{Conclusion}

The effect of freestream turbulence (FST) on the multiscale coherent dynamics in a wind turbine wake was studied using particle image velocimetry measurements. A total of 10 FST cases were studied that had varying turbulence intensity ($T_i$) and integral length scales ($L_v$) much smaller than the turbine diameter. The tip vortices were found to break down closer to the turbine in the presence of FST, resulting in an earlier initiation of wake recovery. \rfC{The onset location of wake recovery ($x_w$) was found to move closer to the turbine as $T_i$ and $L_v$ were increased.} A stronger dependence of $x_w$ on $L_v$ was observed, at least for the low $T_i$ cases considered.

The nature and amplitude of wake meandering at varying FST levels was elucidated. For all the cases, low frequency peaks were observed in the wake that fell in the generally accepted wake meandering frequency band ($0.15 \leq St_D \leq 0.4 $). In the near wake, the power associated with wake meandering was more comparable for all the cases, showing no clear trend with either $T_i$ or $L_v$. This indicated that the process of initiation of wake recovery was the same for all the cases, likely related to a turbine instability \citep{howard2015statistics, foti2016wake, biswas2024effect}. In the far wake however, as the wake meandering power for the no grid case (negligible FST) decayed, that for the cases with non-negligible FST levels sustained, primarily due to the earlier breakdown of the tip vortices. For the highest FST cases a broadening of the wake meandering frequency band was observed \citep{bourhis2025impact}.

The coherent modes associated with the different frequencies were studied by using Optimal Mode Decomposition (OMD). The OMD algorithm identified a larger number of modes in the vicinity of the wake meandering frequency band that were spatially (and spectrally) similar. These modes were grouped/ clustered by using a slightly modified version of mode clustering algorithm originally introduced by \citet{beit2021data}. The algorithm significantly simplified the OMD spectra, particularly in the vicinity of the wake meandering frequency band. As the FST level was increased, the tip vortex related modes became weaker and decayed much closer to the turbine. On the other hand, the wake meandering modes for the cases with FST showed a larger spanwise width.

The energy exchanges to/from the modes were studied using the coherent kinetic energy budget equation derived by \citet{baj2017interscale}. Even with FST, the wake meandering modes were found to be primarily energised by the mean flow, although there were non-negligible contributions from the diffusion term and triadic interactions. In the presence of FST, the earlier breakdown of the tip vortices allowed the wake meandering modes to better utilize the mean velocity gradient, yielding a higher energy transfer from the mean flow. With FST, the pattern of triadic energy exchanges in the tip vortex system remained almost identical to that reported with negligible FST \citep{biswas2024energy}. However, the magnitudes of the energy exchanges were drastically reduced for elevated FST levels. New secondary frequencies were reported for the first time that were produced through non-linear triadic interaction between the tip vortices and wake meandering induced by FST, similarly to our previous studies on multiscale wakes \citep{baj2017interscale, biswas2022energy}.\\

\rfA{Our study confirms that large-scale turbulence ($L_v>D$) is not compulsory to have energetic modes in the meandering frequency band as assumed in the dynamic wake meandering model (DWM). When wake meandering is primarily driven by the turbine, even small scale turbulence can significantly alter wake meandering dynamics, indirectly through breaking down the shield of the tip vortices}. \rfD{The breakdown of the tip vortices introduced several new frequencies in the tip shear layer making it more broadband and turbulent like. The interaction between the shear layer and the wake meandering modes after the tip vortices have broken down is not understood properly. In addition, this work considered the wake only up to a distance of $x=5D$. Future work should conduct a similar analysis looking at the energy exchanges associated wake meandering further downstream. A pertinent question to answer would be: can wake meandering turn into a completely stochastic process further downstream although its initiation could be related to an intrinsic instability of the turbine? In the far wake, possible links could be found between wake meandering and meandering observed in other flows \citep{bailey2018experimental, bolle2024linear}.    }


\section*{\rfB{Apendix 1: Uncertainties in the estimation of the integral
length scales}}
\label{app1}

\rfB{Several factors contribute to the uncertainties in the estimation of the integral length scales. Firstly the correlation functions themselves may depend on the length of the time series used. To check this, we evaluated the $R_{12}^{'}$ for data truncated in time. Figure \ref{fig:R12_convergence} here shows the variation of $R_{12}^{'}$ at the hub height for two different cases, 2c and 4b with different data length (here $N=5456$ is the total data length in time). $R_{12}^{'}$ is found to be generally well converged when the full data is used. The uncertainties in the estimation of the integral length scales were introduced mainly due to the inherent inhomogeneity of the flow cases and due some extent to the use of a fitting function (for cases where it did not fall off to zero).}

 \begin{figure}
  \centerline{
  \includegraphics[clip = true, trim = 0 0 0 0 ,width= 1.1\textwidth]{figures/R12_convergence_case_2c_case_4b_new_grid_experiments.png}  }
 \caption{ Effect of the data length on the estimation of $R_{12}^{'}$ for the cases 2c and 4b. }
\label{fig:R12_convergence}
\end{figure}

\section*{\rfB{Appendix 2: OMD based mode clustering}}
\label{app2}

\rfB{Dynamic Mode Decomposition (DMD) assumes that there is a linear dependency between two successive snapshots ($\boldsymbol{x_{n}}, \boldsymbol{x_{n+1}}$) of the flow which remains constant over time. This can be written as:}

\renewcommand{\theequation}{A2.\arabic{equation}}

\rfB{
\begin{equation}
    \boldsymbol{x_{n+1}} = \boldsymbol{Ax_{n}}, \hspace{1cm} \boldsymbol{A} \in \mathbb{R}^{p\times p}
\end{equation}}

\rfB{Here $p$ denotes the length of a snapshot (the number of PIV vectors). DMD finds $\boldsymbol{A}$ by solving the minimization problem: }

\rfB{
\begin{equation}
    \min \Big \Vert [\boldsymbol{v_2},...\boldsymbol{v_N}] - \boldsymbol{A}[\boldsymbol{v_1},...\boldsymbol{v_{N-1}}] \Big \Vert^2
\end{equation}}

\rfB{
Here, $N$ is the number of PIV snapshots. In Optimal Mode Decomposition (OMD), $\boldsymbol{A}$ is represented in a rank constrained form ($\boldsymbol{LML^T}$) and a two-way optimization is performed (with optimization variables $\boldsymbol{L}$ and $\boldsymbol{M}$):} 

\rfB{
\begin{equation}
    \min \Big \Vert [\boldsymbol{v_2},...\boldsymbol{v_N}] - \boldsymbol{LML^T}[\boldsymbol{v_1},...\boldsymbol{v_{N-1}}] \Big \Vert^2
\end{equation}}

\rfB{Here, $\boldsymbol{L}$ represents the low order subspace of the flow field and $\boldsymbol{M}$ represents the dynamics of $\boldsymbol{L}$. \citet{wynn2013optimal} showed that in DMD, $\boldsymbol{L}$ was taken as the POD subspace of the flow. Thereafter, the best dynamical representation ($\boldsymbol{M}$) of the flow on that subspace was obtained. \citet{wynn2013optimal} argued that the POD subspace is not necessarily the best choice for $\boldsymbol{L}$, hence suggesting a more general, two-way optimization. Finally, after finding the eigenvectors ($\boldsymbol{z_m}$) and eigenvalues (${\lambda_M}$) of $\boldsymbol{M}$, the OMD eigenvectors ($\boldsymbol{\phi_{OMD}}$) and eigenvalues (${\lambda_{OMD}}$) are obtained as: }

\rfB{
\begin{equation}
    \boldsymbol{\phi_{OMD}} = \boldsymbol{Lz_M}, \hspace{1cm} {\lambda_{OMD}} = \log({\lambda_M})/{\Delta} t
\end{equation}}

\rfB{Here, $\Delta t$ corresponds to the time period between two consecutive snapshots. After identifying the OMD modes, similar modes were identified and clustered. The original mode clustering algorithm proposed by \citep{beit2021data} is briefly discussed here along with slight changes that have been introduced. For each OMD mode $\phi_l$, \citet{beit2021data} defined a metric $S_l = [\Re(\phi_l), \Im(\phi_l)]$, where $\Re$ and $\Im$ denote the real and imaginary parts of $\phi_l$ and are stored as column vectors. To assess the similarity between modes the authors used the smallest angle between the subspaces of the modes. The smallest angle between two modes $\phi_l$ and $\phi_m$ was therefore measured by using the statistic $\theta_{lm} = \sin^{-1}\Big({|| S_l - S_m \Big(S_l^TS_m\Big) ||}_2 \Big)$, where $||.||$ denoted the largest singular value of the matrix. For simplicity, the spatial similarity was then represented by $\zeta = \cos(\theta_{lm})$. }


Once the clusters are identified, each of the modes in the cluster are stacked in a matrix (let us denote it as $M$) as column vectors multiplied by a significance parameter or weight $\sigma_l$. If the $k^{th}$ cluster has $n$ modes, then $M_k = [\sigma_1 \phi_1, \sigma_2 \phi_2 ... \sigma_n \phi_n]$. Thereafter, a singular value decomposition of $M_k$ is performed and the first singular vector is used as the mode representing the cluster. \rfD{However, it is worthwhile to note that several aspects of the method are still not strictly defined and deserve further consideration. One of them is how to assign a frequency and growth rate to a clustered mode. Further, which statistic is the best to use as $\sigma$? For example, one could project the snapshots on the OMD modes to obtain the time varying coefficients of the modes. The magnitude of the coefficient at {a specific time} or a norm of the coefficient could be used as $\sigma$.} Here however, it was found that using such a statistic undermined the influence of the lesser damped modes within a cluster. Accordingly, a quantity related to the growth rate ($\Re(\lambda_{OMD})$) of the modes was thought to be a more reasonable measure of $\sigma$ {motivating the use of} the quantity $1/|\Re(\lambda_{OMD})|$. The dominant modes (growth rate closer to zero) are therefore assigned a higher significance value. This method showed faster convergence for the current data set and yielded better converged modes representing a cluster. The modes representing a cluster are also assigned a virtual growth rate and a frequency which is obtained by taking a weighted average of the frequencies and growth rates of the modes forming a cluster with $\sigma$ used as the weight. The whole process is repeated until no two modes are similar.

\begin{figure}
  \centerline{
  \includegraphics[clip = true, trim = 0 0 0 0 ,width= 1\textwidth]{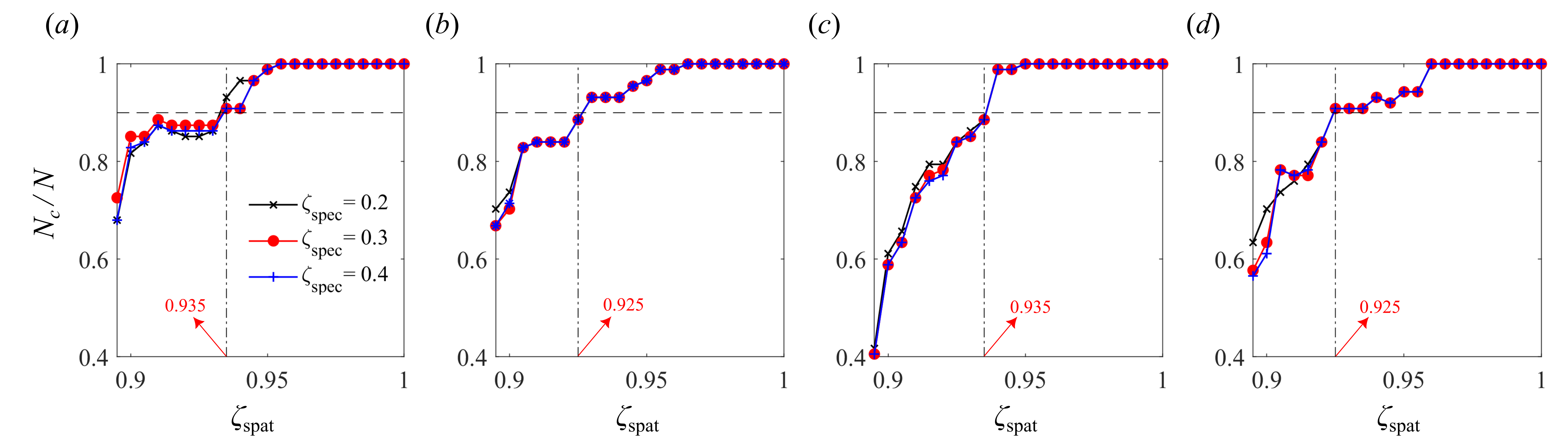}  }
 \caption{Variation of the number of converged modes ($N_c$) with $\zeta_{spat}$ for different $\zeta_{spec}$ for the cases (a) 1, (b) 2c, (c) 3b and (d) 4b.  }
\label{fig:convergence}
\end{figure}

The choice of the cutoff parameters $\zeta_{spat}$ and $\zeta_{spec}$ also involves some subjectivity. The effects of $\zeta_{spat}$ and $\zeta_{spec}$ on the converged number of modes ($N_c$) for the four main cases are shown in figure \ref{fig:convergence}. The rank of OMD was set to 175 (\text{i.e.} the total number of initial modes $N=175$) for all cases following our previous work \citep{biswas2024energy}. No mode reduction is achieved for $\zeta_{spat}\geq 0.96$, while for $\zeta_{spat}\leq 0.9$ the spectra are over simplified leaving too few modes. A different $\zeta_{spat}$ was selected for each case such that $N_c/N \approx 0.9$. For the cases shown in figure \ref{fig:convergence}, the selected cutoffs are shown by the vertical dashed line. The different symbols in figure \ref{fig:convergence} correspond to different values of $\zeta_{spec}$. Note that the mode clustering process does not depend significantly on $\zeta_{spec}$ for $\zeta_{spec} \geq 0.2$. The analysis was therefore continued with $\zeta_{spec}=0.3$ for all the cases.

 \begin{figure}
  \centerline{
  \includegraphics[clip = true, trim = 0 0 0 0 ,width= \textwidth]{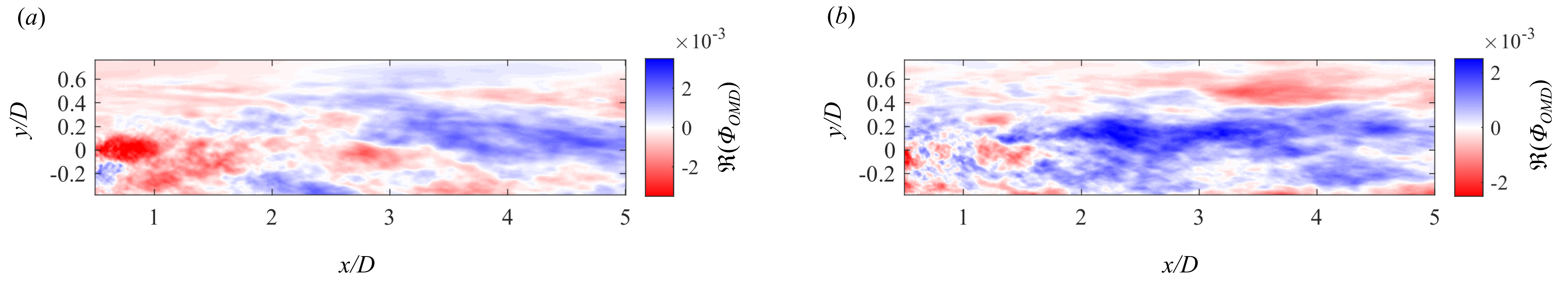}  }
 \caption{ Example low frequency modes for the (a) no grid case and (b) case 4b that are not considered to be a part of the coherent component of the flow {and are likely attributable to an intermittent large-scale phenomenon}. The real part of the OMD modes ($\Re(\phi_{OMD})$) is shown.  }
\label{fig:low_f_struc}
\end{figure}

\section*{Appendix 3: Mode selection and reduced order representation of the flow}
\label{app3}

{Although mode clustering simplifies the OMD spectrum, care must be taken while selecting modes from the reduced spectrum (after performing mode clustering) to represent the coherent component of the flow. The lesser damped modes generally represent a physically meaningful mode, but drawing a cutoff on the growth rate ($\Re(\lambda_{OMD})$) is subjective. Looking at the nature of the spectra, the modes $\Re(\lambda_{OMD}) \gtrsim -0.05$, or in that vicinity, were considered. Before selecting a mode, each mode shape had to be checked manually to see if they show any physically meaningful structure. Sometimes a slightly more damped mode had to be selected since its spatial coherence was more interpretable in a physically meaningful way. Occasionally some modes were not selected despite being less damped than some other selected modes (example figure \ref{fig:iter}(c)). The spatial shape of these modes revealed very large scale smeared structures which were likely due to the intermittent passage of large scale structures having a size smaller than that indicated by the mode shapes. Example of such modes are shown in figure \ref{fig:low_f_struc} for the no grid case and case 4b. These modes are also highlighted by a blue arrow in the spectra shown in figures \ref{fig:iter}(c) and \ref{fig:iter}(l). Furthermore, when a mode was obtained by clustering other modes, the growth rate of the new mode was simply calculated as a weighted average of the growth rates of the modes forming the cluster. This was done simply to place the new mode somewhere in the spectrum and this growth rate need not have any physical meaning. In summary, the shape of the modes involved in clustering was always checked irrespective of their growth rates. }  \\

\noindent \textbf{Funding}. NB gratefully acknowledges funding through the Imperial College London President's Scholarship and the Engineering and Physical Sciences Research Council (EPSRC) through grant EP/T51780X/1. OB gratefully acknowledges funding from EPSRC through grant no. EP/V006436/1.\\

 \noindent \textbf{Declaration of interests}. The authors report no conflict of interest.\\

 \noindent For the purposes of open access, the authors have applied a Creative Commons Attribution (CC BY) licence to any Author Accepted Manuscript (AAM) version arising.

\bibliography{bib}

\end{document}